\documentclass[]{aa}   
\usepackage{graphicx}
\usepackage{txfonts}
\usepackage{enumerate}
\usepackage{natbib}
\usepackage{answers}
\usepackage{subfig}

\begin{document}
\title{Study of underlying particle spectrum during huge X-ray flare of Mkn 421 in April 2013}
\titlerunning{Study of Mkn 421 during huge X-ray flare in April 2013}

\author{Atreyee Sinha$^{1}$, A. Shukla$^{1,3}$, R. Misra$^{2}$, V. R. Chitnis$^{1}$, A. R. Rao$^{1}$, B. S. Acharya$^{1}$}
\authorrunning{Sinha et. al.}

\institute{
\inst{1}~Tata Institute of Fundamental Research, Homi Bhabha Road, Colaba, Mumbai, 400 005, India.\\ 
\inst{2}~Inter-University Center for Astronomy and Astrophysics, Post Bag 4, Ganeshkhind, Pune-411007, India.\\ 
\inst{3}~Now at ETH Zurich, Institute for Particle Physics, Otto-Stern-Weg 5, 8093 Zurich, Switzerland. \\
\email{atreyee@tifr.res.in} \\
}
\abstract
%
{ In April 2013, the nearby (z=0.031) TeV blazar, Mkn 421, showed  one of the largest flares in X-rays since the past decade.}
{ To study all multiwavelength data available during MJD 56392 to 56403, with special emphasis on X-ray data, and understand the underlying particle energy distribution.}
{We study the correlations between the UV and gamma bands with the X-ray band using the z-transformed discrete correlation function. 
We model the underlying particle energy spectrum with a single population of electrons emitting synchrotron radiation, and do a statistical fitting of the simultaneous, time-resolved data from the {\it Swift}-XRT and the {\it Nu}STAR.}
{There was rapid flux variability in the X-ray band, with a minimum doubling timescale of $1.69 \pm 0.13$ hrs. There were no corresponding flares in UV and gamma bands. The variability in UV and gamma rays are relatively modest with $ \sim 8 \% $ and $\sim 16 \% $ respectively, and no significant correlation was found with the X-ray light curve.
The observed X-ray spectrum shows clear curvature which can be fit by a log parabolic spectral form. This is best explained to originate from a log parabolic electron
spectrum. However, a broken power law or a power law with an exponentially falling electron distribution cannot be ruled out either. 
Moreover, the excellent broadband spectrum from $0.3-79$ keV allows us to make predictions of the UV flux. We find that this prediction is compatible with the observed flux during the low state in X-rays. However, during the X-ray flares, depending on the adopted model, the predicted flux is a factor of $2-50$ smaller than the observed one. This suggests that the X-ray flares are plausibly caused by a separate population which does not contribute significantly to the radiation at lower energies. Alternatively, the underlying particle spectrum can be much more complex than the ones explored in this work.}
{}
\keywords{BL Lacertae objects: individual (Mkn 421)- galaxies: active - X-rays: galaxies - radiation mechanisms: non-thermal}

\maketitle


\section[sect:Intro]{Introduction}

According to the unification scheme of Active Galactic Nuclei (AGNs) by \cite{UrryPadovani}, 
blazars are a subclass of AGNs with a relativistic jet aligned close to the line of sight. 
Blazars are further subdivided into BL~Lacs and FSRQs where BL~Lacs are characterized by the absence of (or very weak) emission lines. They show high optical polarization, intense and highly variable non-thermal radiation throughout
the entire electromagnetic spectra in time scales extending from minutes to years, apparent super-luminal motion in
radio maps, large Doppler factors and beaming effects. The broadband spectral energy distribution (SED) of blazars is characterized by two peaks, one in the IR - X-ray regime, and the second one in $\gamma$-ray regime. According to the location of the first peak, BL Lacs are further classified into Low energy peaked BL~Lacs (LBLs) and High energy peaked BL~Lacs (HBLs) \citep{PadGio}. Both leptonic and hadronic models have been used to explain the broadband SED with
varying degrees of success. The origin of the low energy component is well established to be caused by synchrotron emission from
relativistic electrons gyrating in the magnetic field of the jet. However the physical mechanisms responsible for the high energy emission are
still under debate. It can be produced either via inverse Compton (IC) scattering of low frequency photons by the same
electrons responsible for the synchrotron emission (leptonic models), or via hadronic processes initiated by relativistic
protons, neutral and charged pion decays or muon cascades (hadronic models). The seed photons for IC in leptonic
models can be either the synchrotron photons itself (Synchrotron Self Compton, SSC) or from external sources such
as the Broad Line Region (BLR), the accretion disc, the cosmic microwave background, etc (External Compton, EC). For a comprehensive review of
these mechanisms, see \cite{bottcher2007}. 

Mkn 421 is the closest ($z=0.031$) and the most well studied TeV blazar. It was also the first detected extragalactic TeV source \citep{punch}, and one of the brightest BL~Lac objects seen in the UV and X-ray bands. It is a HBL, with the synchrotron spectrum peaking in the X-ray regime. Moreover, the X-ray emission is  known to be highly correlated with the TeV emission  \citep{multi421,Katarzynski_hbl,Qian}, but like other HBLs, shows moderate correlation with the GeV emission \citep{Xray-GeV_corr}.
It is highly variable and has been well studied during its flaring episodes by several authors \citep{MAGIC421_2008,Amit421,MAXI421,Mkn421_2000,MAGIC421_2006,MAGIC421_outburst,Suzaku421_2006,Swift421_2006,multi421_2005-2006,Integral421,421_march2001,MAGIC421,XMM421_2003}. 
A detailed study of its quiescent state emission has been performed by \citet{fermi421} with the most well sampled SED till date. 

During April 2013, Mkn 421 underwent one of the largest X-ray flares ever recorded in the past decade \citep{Pian421}.
The source was simultaneously observed by {\it Swift} and {\it Nu}STAR during this flaring episode, and we use  these observations to study the spectral variations. To strengthen our study further, we supplement the X-ray information with other multiwavelength observations available.

The main focus of our work being the joint spectral fitting between the Nuclear Spectroscopic Telescope Array {\it Nu}STAR the {\it Swift}-XRT telescopes, and using this, we investigate the underlying particle energy spectrum.
The high photon statistics during the flare, coupled with the excellent spectral response of {\it Swift}-XRT and {\it Nu}STAR gives us a rich X-ray spectrum from $0.3 - 79 keV$ to explore.
A study of the quiescent state of Mkn 421 using {\it Nu}STAR data was performed by \citet{nustar421}. In Section \ref{sect:Analysis} we describe the data reduction techniques from the various instruments. Section \ref{sect:Multi} lists the multiwavelength temporal results, while the X-ray spectral modelling is described in Section \ref{sect:spectra}. We discuss the implications of the results in Section \ref{sect:Disc}.

\section[sect:Analysis]{Multiwavelength Observations and Data Analysis}\label{sect:Analysis}

The huge X-ray flare of Mkn 421 during April 2013 ( 2013 April 10 to 21; MJD 56392$-$56403)  is simultaneously observed by the {\it Nu}STAR and {\it Swift} X-ray and UV telescopes. The $\gamma$-ray behavior of this source during this flare was obtained by analyzing {\it Fermi}-LAT observations. In addition, we also include the X-ray observations by MAXI and optical observations by SPOL for the present study. The analysis procedures of these observations are described below.

\subsection{{\it Fermi}-Large Area Telescope Observations}\label{subsec:fermi}

The {\it Fermi}-LAT data used in this work were collected covering the period of the X-ray outburst (MJD 56390$-$56403).
The standard data analysis procedure as mentioned in the {\it Fermi}-LAT documentation\footnote{http://fermi.gsfc.nasa.gov/ssc/data/analysis/documentation/}
was employed. Events belonging to the energy range 0.2$-$300 GeV and SOURCE class were used. To select good time intervals, 
a filter ``\texttt{DATA$\_$QUAL$>$0}'' \&\& ``\texttt{LAT$\_$CONFIG==1}'' was chosen and only events with less than 105$^{\circ}$ zenith angle were selected to avoid contamination from the Earth limb $\gamma$-rays. 
The galactic diffuse emission component gll\_iem\_v05\_rev1.fits and an isotropic component iso\_source\_v05\_rev1.txt were used as the background models. The unbinned likelihood method included in the pylikelihood library of {\tt Science Tools (v9r33p0)} and the post-launch instrument response functions P7REP\_SOURCE\_V15 were used for the analysis. All the sources lying within 10$^{\circ}$ region of interest (ROI) centered at the position of Mkn 421 and 
defined in the second {\it Fermi}-LAT catalog \citep{2FGL}, were included in the xml file. All the parameters except the scaling factor of the sources within the ROI are allowed to vary during the likelihood fitting. For sources between 10$^{\circ}$ to 20$^{\circ}$ from the centre, all parameters were kept frozen at the default values. The source was modelled by a power law as in the 2FGL catalog.

\subsection{{\it Nu}STAR Observations}\label{subsec:nustar}

   {\it Nu}STAR \citep{Nustar_instru} features the first focussing X-ray telescope to extend high sensitivity beyond 10 keV. 
There were 11 {\it Nu}STAR pointings between the aforementioned dates, the details of which are given in Table \ref{nupoint}. 
The {\it Nu}STAR data were processed with the NuSTARDAS software package v.1.4.1 available within HEASOFT package (6.16). The latest CALDB (v.20140414) was used. 
After running {\tt nupipeline v.0.4.3} on each observation, {\tt nuproducts v.0.2.8} was used to obtain the lightcurves and spectra. 
Circular regions of 12 pixels centered on Mkn 421, and of 40 pixels centered at 165.96, 38.17  were used as source and background regions respectively. The spectra from the two detectors A and B were combined using {\tt addascaspec}, and then grouped (using the tool {\tt grppha v.3.0.1}) to ensure a minimum of 30 counts in each bin. To get strict simultaneity with {\it Swift}-XRT observations, observation id 60002023025 was broken into 4 parts 56393.15591890  to 56393.29538714, 56393.29538714 to 56393.91093765, 56393.91093765 to 56393.96788711 and 56393.96788711 to  56394.37822500 MJD.

\subsection{{\it Swift} Observations}\label{subsec:swift}

There were 15 {\it Swift} pointings between the aforementioned dates, the details of which are given in Table \ref{swiftpoint}. Publicly available daily binned source counts were taken from the {\it Swift}-BAT webpage\footnote{http://swift.gsfc.nasa.gov/results/bs70mon/SWIFT\_J1104.4p3812}.

The XRT data \citep{XRT_instru} were processed with the XRTDAS software package (v.3.0.0) available within HEASOFT package (6.16). Event files were cleaned and calibrated using standard procedures ({\tt xrtpipeline v.0.13.0}), and {\tt xrtproducts v.0.4.2} was used to obtain the lightcurves and spectra.
 Standard grade selections of $0-12$ in the Windowed Timing (WT) mode are used. 
Circular regions of 20 pixels centered on Mkn 421 (at 166.113 and Dec 38.208) and of 40 pixels centered at 166.15, 38.17 were used as source and background regions respectively.
For the observations affected by pileup (counts $> 100 c/s$) \citep{XRT_pileup}, an annular region of inner radius 2 pixels and outer 20 pixels was taken as the source region. 
The lightcurves were finally corrected for telescope vignetting and PSF losses with the tool {\tt xrtlccorr v.0.3.8}. The spectra were grouped to ensure a minimum of 30 counts in each bin by using the tool {\tt grppha v.3.0.1}.

{\it Swift}-UVOT \citep{UVOT_instru} operated in imaging mode during this period, 
and for most of the observations, cycled through the UV filters UW1, UW2 and UM2. 
The tool {\tt uvotsource v.3.3} was used to extract the fluxes from each of the images using aperture photometry. 
The observed magnitudes were corrected for galactic extinction ($E_{B-V}=0.019$ mag) using the dust maps of \cite{schelgel} and converted to flux units using the zero point magnitudes and conversion factors of \citet{UVOT_conv}. The tool {\tt flx2xsp v.2.1} was used to convert the fluxes to pha files for use in XSPEC. 

\subsection{Other Multiwavelength data}\label{subsec:other}
Publicly available daily binned source counts were plotted for MAXI \footnote{http://maxi.riken.jp/}. As a part of the {\it Fermi} multiwavelength support program,
 the SPOL CCD Imaging/Spectropolarimeter at Steward Observatory at the University of Arizona \citep{CCD-SPOL} 
 regularly observes Mkn 421. The publicly available optical V-band photometric and linear polarization data were downloaded from their website \footnote{http://james.as.arizona.edu/$\sim$psmith/Fermi/}.

 \section[sect:Multi]{Multiwavelength Temporal Study}\label{sect:Multi}

\begin{figure*}
	\centering
	\includegraphics[scale=2.3]{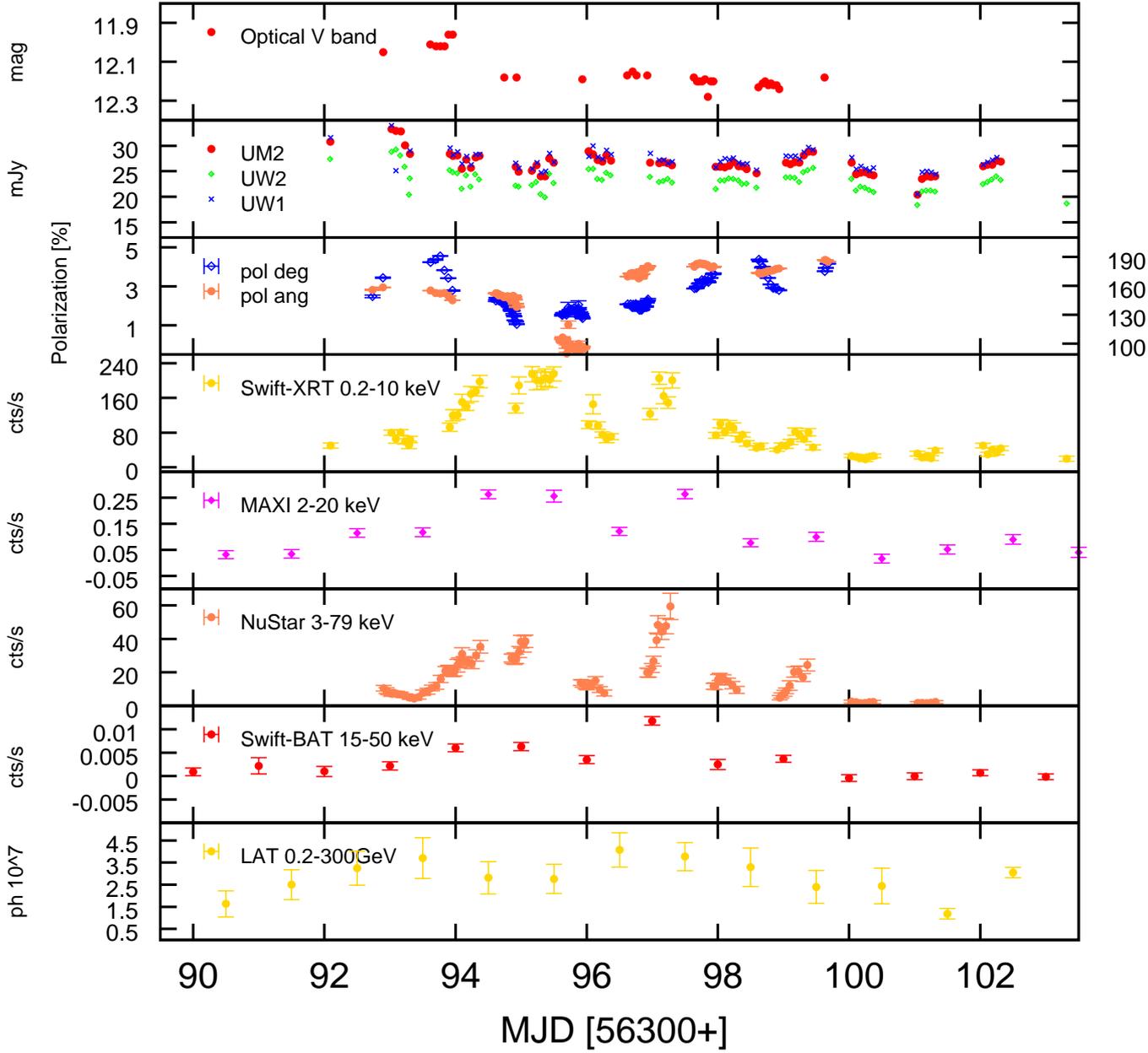}
	\caption{Multiwavelength lightcurve during MJD 56390 to 56403 showing :
	Panel 1: V band magnitude; Panel 2: UV flux in mJy; Panel 3: Degree and angle of optical polarization;
Panel 4: {\it Swift}-XRT flux in counts/sec; Panel 5: MAXI flux in counts/sec; Panel 6: {\it Nu}STAR flux in counts/sec; Panel 7: {\it Swift}-BAT flux in counts/sec; Panel 8: {\it Fermi}-LAT flux in $ph/cm^2/sec$. One {\it Swift}-XRT (or {\it Nu}STAR) point is plotted for each snapshot.  }
	\label{fig:lc_all}
\end{figure*}

The Multiwavelength lightcurve during the 10 days period, from optical to gamma ray energies, along with the optical polarization measurements, is plotted in Fig. \ref{fig:lc_all}. While there were two huge flares in X-rays (on MJD 56395 and 56397), where the flux went up by a factor of 10, the fluxes in the other bands were not very variable on the timescale of days. We compute the z-transformed discrete correlation using a freely available Fortran 90 code with the details of the method employed described in \citet{zdcf}. 
We find no lag between the soft ($0.3-10$ keV, {\it Swift}-XRT)  and the hard X-ray ($3.0-79$ keV {\it Nu}STAR) bands. There is no correlation seen between the UV flux and the X-ray flux, ($zdcf_{max}=0.62 \pm 2.3$, at a lag of $2.2$ days). Also, while the UV flux does not show correlation with the optical polarization ($zdcf_{max}=0.61 \pm 1.1$, at a lag of $2.4$ days), the X-ray flux shows a tighter correlation ($zdcf_{max}=0.81 \pm 0.6$, at a lag of $3.5$ days) with the latter.
There was also a large change in the angle of polarization during the two X-ray flares.  

\begin{figure*}
	\centering
	\includegraphics[scale=1.2]{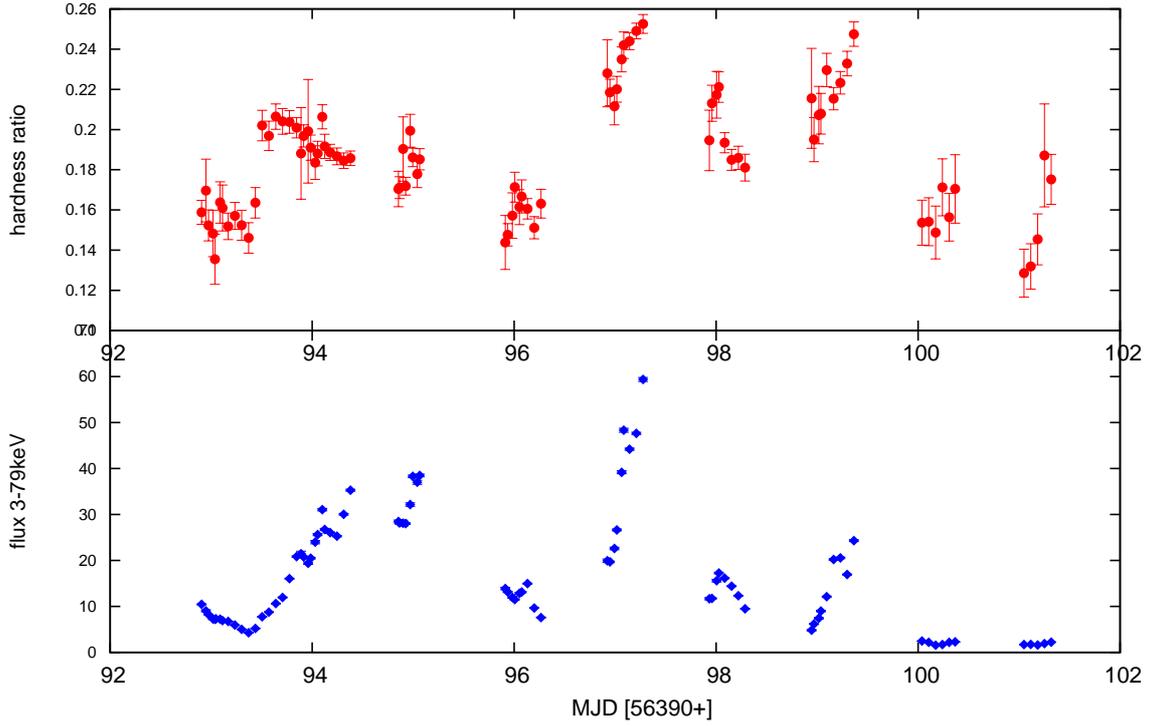}
	\caption{Plot of hardness ratio as observed by {\it Nu}STAR FPMA. The upper panel shows the hardness ratio, and the lower one, the flux(counts/sec). There is a trend of spectral hardening with flux.}
	\label{fig:hardness}
\end{figure*}

The hardness ratios (computed here as the ratio between the 10 $-$ 79 keV count rate and the 3 $-$ 10 keV count rate \citep{harddefn})  are plotted in Fig. \ref{fig:hardness}. A trend of spectral hardening with increasing flux (Spearman rank correlation, rs=$0.58$, $p=5.0\cdot10^{-8}$), is observed and the same is often reported for this source \citep[eg:][]{nustar421,XMM421}. Moreover, the correlation is much tighter during the rising part of the 2 flares (rs=$0.92$, $p=4.0\cdot10^{-5}$). These interesting features advocate us to perform a more detailed spectral study, which we describe in Section \ref{sect:spectra}.

The fractional variability amplitude parameter $F_{var}$ \citep{Vaughan,varsha_var}, computed on daily timescales, is used to quantify the multi-wavelength variability. It is calculated as

\begin{equation} 
	F_{var}=\sqrt{\frac{S^2-\sigma^2_{err}}{\bar{x}^2}}
 \end{equation} 
\noindent
where $\sigma^2_{err}$ is the mean square error, $\bar{x}$ the unweighted sample mean, and $S^2$ the sample variance. The error on $F_{var}$ is given as 
\begin{equation}
	\sigma_{F_{var}}= \sqrt{ \left( \sqrt{\frac{1}{2N}}\cdot\frac{\sigma^2_{err}}{\bar{x}^2 F_{var}} \right)^2 +  \left( \sqrt{\frac{\sigma^2_{err}}{N}}\cdot\frac{1}{\bar{x}} \right)^2}
\end{equation}
\noindent
Here, $N$ is the number of points.

The variability amplitude is the maximum for the X-ray bands
($F_{var} \sim 0.75 \pm 0.10 $), and significantly lower for the UV bands  
($F_{var} \sim 0.08 \pm 0.02 $), suggesting that the emission may probably arise from different components in the two bands. The variability in the GeV range is also small   ($F_{var} \sim 0.16 \pm 0.07 $).  This goes against the general trend found in blazars that $F_{var}$ is the maximum for the $\gamma$-ray band and decreases with frequency \citep{Zhang2155,Vaidehi3C279}. 

We scan the {\it Swift}-XRT and the {\it Nu}STAR lightcurves for the shortest flux doubling timescale using the following equation \citep{fluxdoubling}
\begin{equation}\label{eq:flux_double}
F(t) = F(t_0).2^{(t-t_0)/\tau}
\end{equation}
where $F(t)$ and $F(t_0)$ are the fluxes at time $t$ and $t_0$ respectively, and $\tau$ is the characteristic doubling/halving time scale. 
The fastest observed variability in the {\it Nu}STAR band is $1.69 \pm 0.08$ hours between MJD 97.06438 to 97.08573. This is comparable to 
the very fast variability observed in this source with the {\it Beppo}-SAX \citep{BeppoSAX-fossati}.

The above study is performed only for those periods where the flux difference is significant, at least at the $3 \sigma$ level.

\section[sect:spectra]{X-ray spectral analysis}\label{sect:spectra}

\begin{figure*}
	\centering
	\includegraphics[scale=1.2]{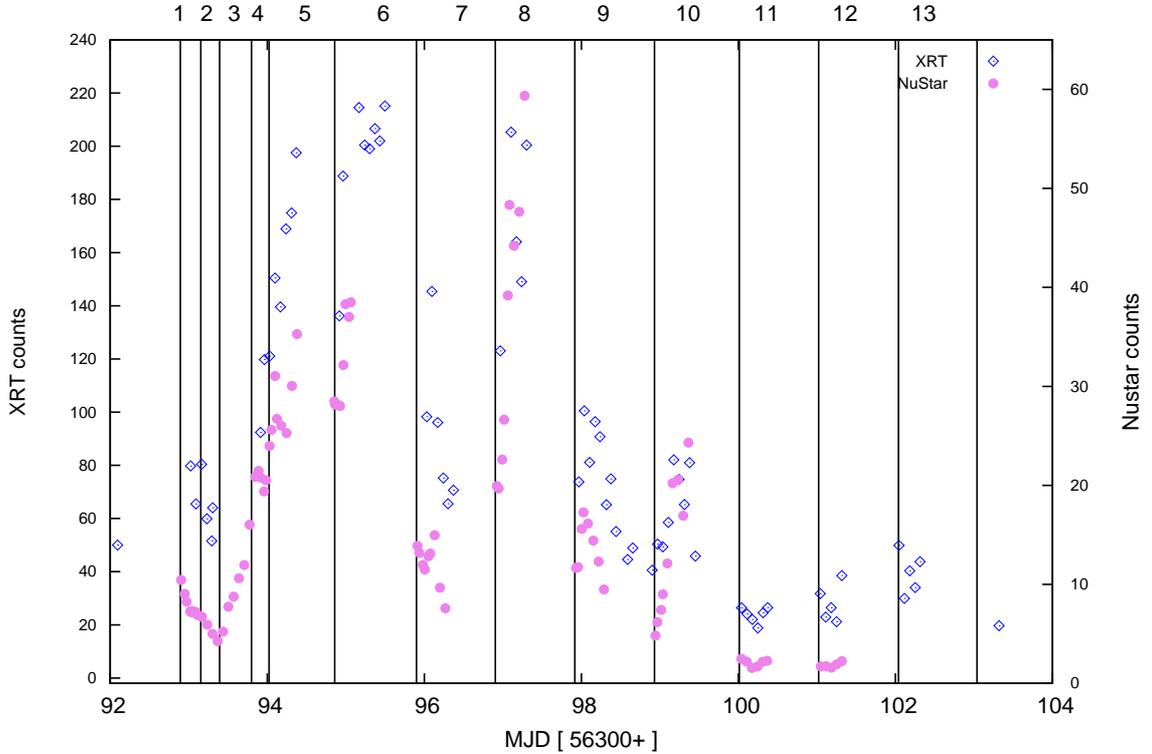}
	\caption{The 13 time bins for which spectra have been extracted.}
	\label{state}
\end{figure*}

The close simultaneity between  {\it Swift}-XRT and {\it Nu}STAR observations allows us to do a joint spectral fitting using {\tt XSPEC package v.12.8.2}. The time periods which have been fitted together are shown in Fig. \ref{state}. The bin widths are selected as one bin per {\it Swift}-XRT observation (except for Obs. id. 00035014062 and 00035014069, which last only for a few minutes) leading to 13 time bins which we denote as {\bf f1 - f13}. The state {\bf f3} has no {\it Swift}-XRT data, whereas the state {\bf f13} has no {\it Nu}STAR data. Again in Table \ref{jointid} we list the {\it Swift}-XRT and corresponding NuSTAR data which have been combined.

While fitting the broadband X-ray spectrum ($0.3-79$ keV), the XRT and the NuSTAR spectral parameters were tied to each other, except the relative normalization between the two instruments. To correct for the line of sight absorption of soft X-rays due to the interstellar gas, the neutral hydrogen column density was fixed at $N_{H}= 1.92 \times 10^{20} cm^{-2} $ \citep{LAB}.

\subsection[sect:photspec]{Fitting the photon spectrum}\label{sect:photspec}

It is known that the X-ray spectrum of Mkn 421 shows significant curvature \citep{fossati421,massaro_lp1}, and consistently we also noted that the data cannot be fitted satisfactorily by a simple power law. On the other hand, a power law with an exponential cutoff gives a much steeper curvature than observed, yielding to unacceptable fits. A sharp broken power law also gives large $\chi^2$ values in most cases, which suggests a smooth intrinsic curvature in the spectrum. So, following \citep{massaro_lp1,swift_lp} we fit the observed spectrum with a log parabola given by
\begin{equation}
	 dN/dE = K(E/E_b)^{-\alpha_s -\beta_s log(E/E_{b,s})}
 \end{equation}
\noindent
 where $\alpha_s$ gives the spectral index at $E_{b,s}$. 
 The point of maximum curvature, $E_{p,s}$ is given by 
\begin{equation}
	\label{lp_phot}
	E_{p,s} = E_{b,s} 10^{(2-\alpha_s)/2\beta_s}
\end{equation}
\noindent
During fitting, $E_{b,s}$ is fixed at $1keV$. In Table \ref{jointid}, we give the resulting reduced $\chi^2$s for the case of the broken power-law and the log parabola models; while in Table \ref{lp_par}, we give the fit parameters corresponding to the latter model.

\begin{figure*}
	\subfloat[$E_p$ vs $\alpha$]{\includegraphics[scale=0.6]{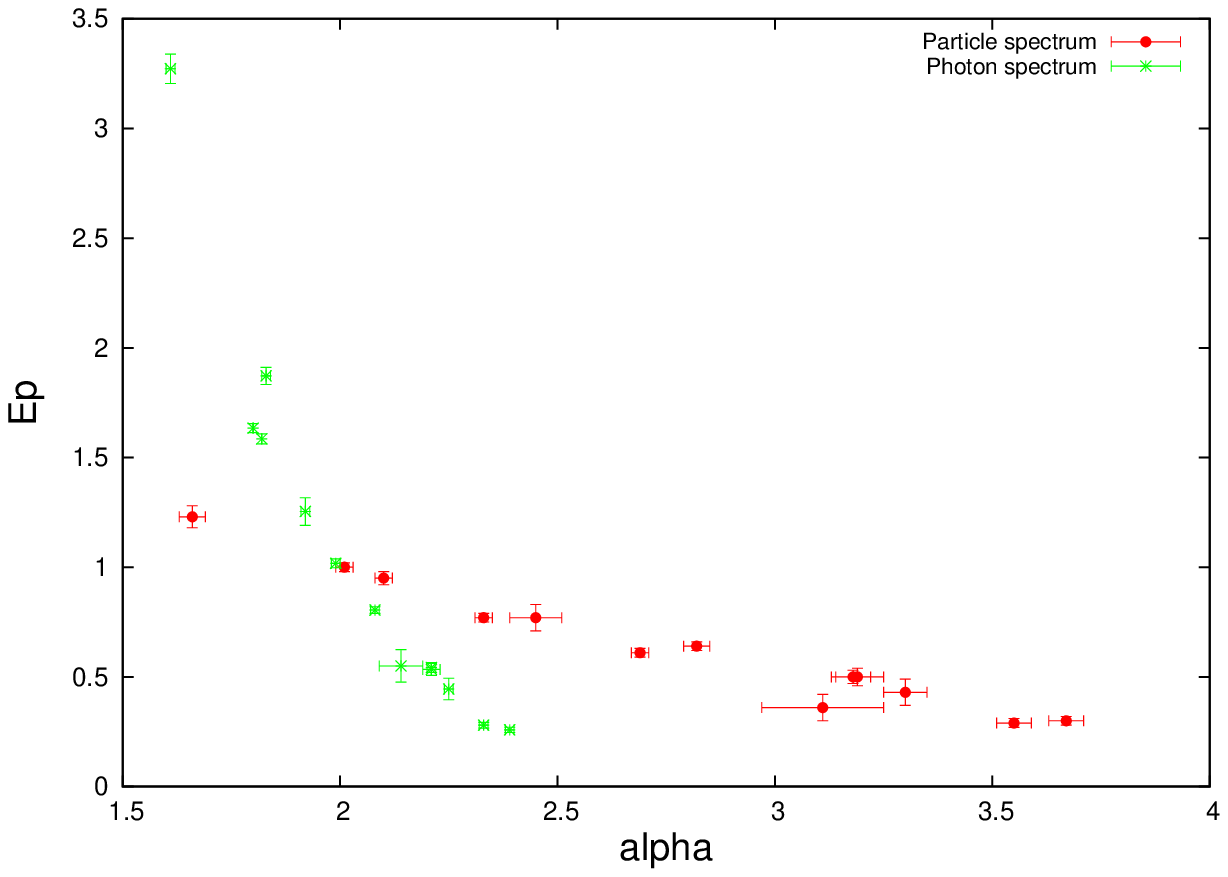}}
	\qquad
	\subfloat[$\beta$ vs $\alpha$]{\includegraphics[scale=0.6]{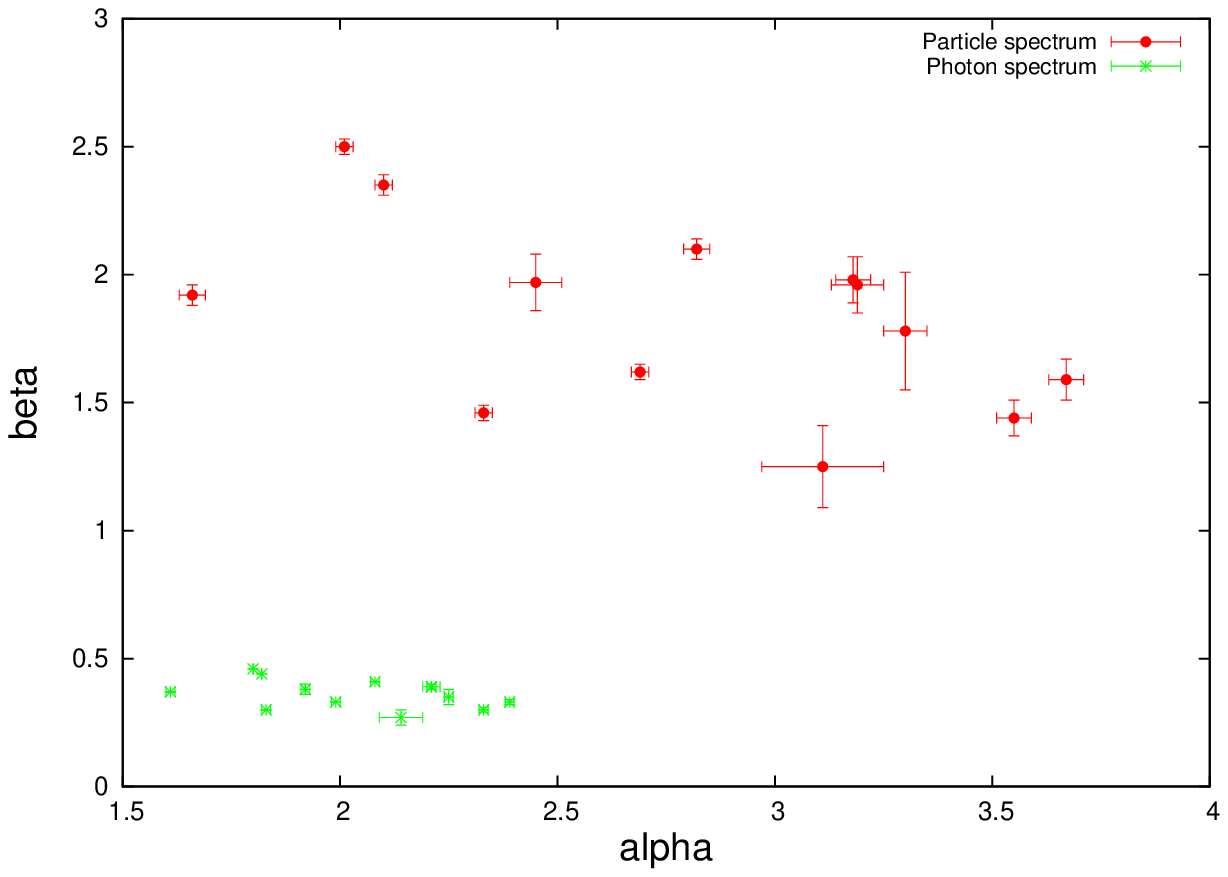}}
	\qquad
	\subfloat[$E_p$ vs $\beta$]{\includegraphics[scale=0.6]{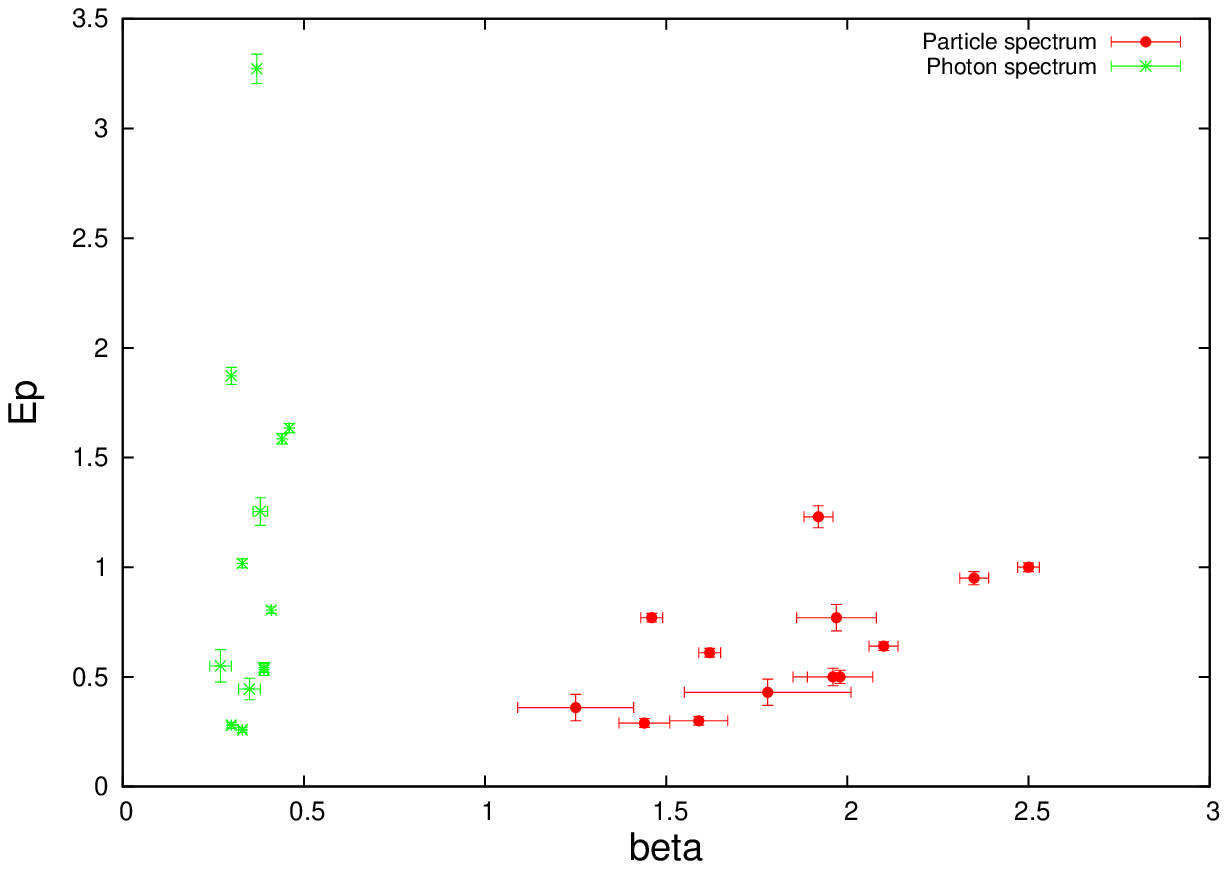}}
	\qquad
	\subfloat[$E_p$ vs flux]{\includegraphics[scale=0.6]{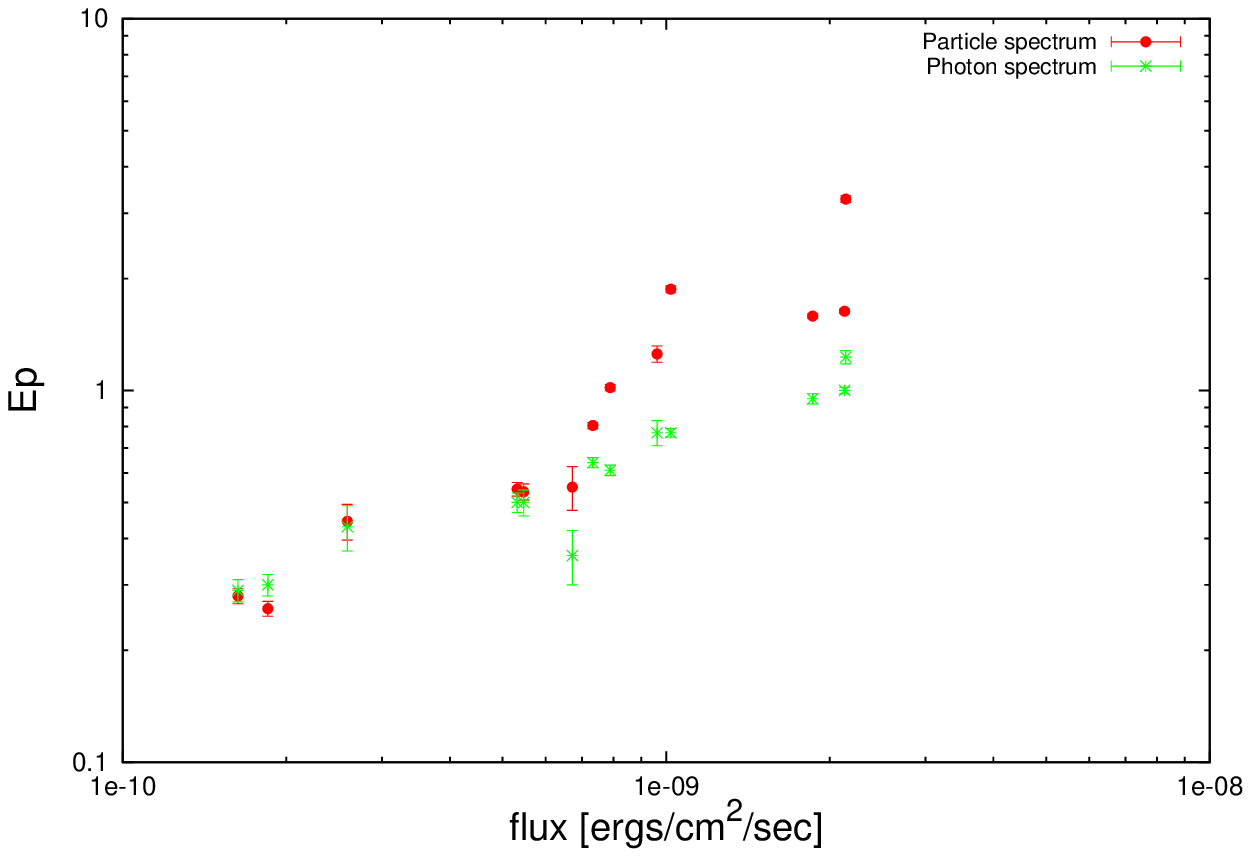}}
	\caption{Cross plot of parameters for the logparabolic photon (green stars) and particle (red circles) spectrum respectively. The peak curvature $E_p$ clearly anti-correlates with $alpha$, and strongly correlates with the flux. However, there is no correlation between $E_p$ and the curvature parameter $\beta$, or between $\beta$ and $\alpha$.}
	\label{logpar_plt}
\end{figure*}

Interestingly we noticed a strong anti-correlation between $\alpha_s$ and flux ($rs = -0.98$, $p < 2.2 \cdot 10^{-16}$), and a strong correlation between flux and $E_{p,s}$ ($rs = 0.97$, $p < 2.2 \cdot 10^{-16}$) which implies that during flares, the spectral index at $1keV$ hardens and the peak of the spectrum shifts to higher energies. This behaviour of the source has often been reported \citep{massaro_lp1,lp_many}. In addition, we did not see any correlation between $\alpha_s$ and $\beta_s$ ($rs = -0.35$, $p = 0.24$), which was seen by \cite{massaro_lp1}. We also noticed that there is no correlation observed between the curvature parameter $\beta$ and the peak of the curvature $E_p$ ($rs = 0.63$, $p = 0.02$). These cross plots are shown in Figure \ref{logpar_plt}. 

\subsection[sect:partspec]{Emitting Particle Distribution}

The excellent spectral resolution of {\it Nu}STAR gives us an unprecedented view of the high energy X-ray behavior beyond $20keV$. Coupled with {\it Swift}-XRT, we have, for the first time, an uninterrupted, well resolved spectrum from $0.3 - 79$ keV. This allows us to go beyond only fitting the photon spectrum with various spectral forms. Rather, in this work, we try to study the underlying particle distributions which give rise to the observed photon spectrum.

\begin{figure*}
	\centering
	\includegraphics[scale=0.9]{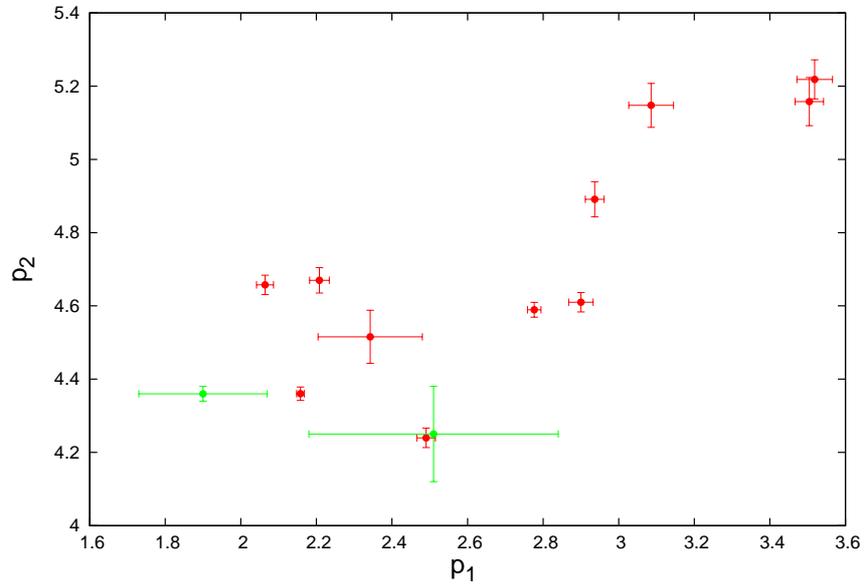}
	\caption{Cross plot of the 2 indices of the broken power law particle spectrum. The green points are not from combined spectrum, but are 
	{\it Nu}STAR only and {\it Swift}-XRT only (see text).} 
	\label{p1p2}
\end{figure*}

We consider the case where X-ray emission arise from a relativistic distribution of electrons emitting synchrotron radiation. The electrons are confined within a spherical zone of radius $R$ filled with a tangled magnetic field $B$. Due to relativistic motion of the jet, the radiation is boosted along our line of sight by a Doppler factor $\delta$. A good sampling of the entire SED from radio to $\gamma$-rays allows one to perform a reasonable estimation of these physical parameters \citep{tavKN}. For the case of synchrotron emission alone, $R$, $B$ and $\delta$ will only decide the spectral normalisation. On the other hand, the shape of the observed spectrum is determined by the corresponding form of the underlying particle spectrum. To obtain further insight into the emitting particle distribution, we developed synchrotron emission models with different particle distribution and incorporated them into XSPEC spectral fitting software. Particularly for this study, we consider the following particle distributions: 
\begin{enumerate}[(i)]

\item Simple power law (SPL): In this case, we assume the electron distribution to be a simple power law with a sharp high energy cutoff, given by

\begin{equation}
\label{eq:simp}
	N(\gamma) d\gamma =K \gamma^{-p}d\gamma,\mbox {~$\gamma<\gamma_{max}$~} \\
\end{equation}
\noindent

Here $\gamma mc^2$ is the energy of the emitting electron, $p$ is the particle spectral index, $K$ is the normalization and $\gamma_{max} mc^2$ the cut-off energy. Among these parameters, $p$ and $\gamma_{max}$ are chosen as the free parameters. 

\item Cutoff power law (CPL) : Here the underlying particle distribution is assumed to be a power law with index $p$ and an exponential cut-off above energy $\gamma_{m} mc^2$ given by

\begin{equation}
\label{eq:cutoff}
N(\gamma) d\gamma =K \gamma^{-p} exp(\frac{\gamma}{\gamma_m}) d\gamma \\
\end{equation}
\noindent
For this distribution, $p$ and $\gamma_{m}$ are chosen as the free parameters.

\item Broken power law (BPL): The particle distribution in this case is described by a broken power law with indices $p_1$ and $p_2$ with a break at energy $\gamma_b$ given by 
\begin{equation}
\label{eq:broken}
N (\gamma) d\gamma =\left\{
\begin{array}{ll}
K \gamma^{-p_1}d\gamma,&\mbox {~$\gamma_{min}<\gamma<\gamma_b$~} \\
K \gamma^{(p_2-p_1)}_b \gamma^{-p_2}d\gamma,&\mbox {~$\gamma_b<\gamma<\gamma_{max}$~}
\end{array}
\right.
\end{equation}
\noindent
Here, $p_1$, $p_2$ and $\gamma_b$ are chosen as the free parameters.

\item Log parabola (LP): For this case, the particle distribution is chosen to be a log parabola, given by 

\begin{equation}
	\label{lp-part}
	N (\gamma) d\gamma = K (\gamma/\gamma_b)^{-\alpha_p -\beta_p log(\gamma/\gamma_b)} d\gamma
 \end{equation}
\noindent

 with $\alpha_p$ and $\beta_p$ chosen as the free parameters. 
\end{enumerate}
 We fitted the observed combined X-ray spectrum from {\it Swift}-XRT and {\it Nu}STAR with the synchrotron emission due to these different particle distributions as given above. A poor fit statistic with large reduced $\chi^2$ is encountered for the case of SPL since it fails to reproduce the smooth curvature seen in almost all spectral states ($f1 - f13$). For CPL, the statistics improved for many states (Table \ref{jointid}) with lowest reduced $\chi^2$ of 1.01 during state {\bf f3}; whereas, the largest reduced $\chi^2$ is 1.41 for the state {\bf f6}. The fit statistics improved considerably for many states for the case of BPL except for {\bf f6} and {\bf f8} corresponding to peak X-ray flare (Table \ref{jointid}). In Figure \ref{p1p2}, we show the cross plot distribution of the power law indices, $p_1$ and $p_2$, of BPL model during different spectral states. The index $p_1$ is poorly constrained for the state {\bf f3} due to the absence of {\it Swift}-XRT observation during this period; whereas, for state {\bf f11} neither $p_1$ nor $p_2$ are well constrained due to the absence of {\it Nu}STAR observation. However, for most of the states, a strong correlation between $p_1$ and $p_2$ is observed ($rs=0.79$ and $p=0.0018$). Among all these particle distributions, the best statistics is obtained for the case of the LP model with the reduced $\chi^2$ decreased considerably during the flaring states, {\bf f5}, {\bf f6} and {\bf f8} (Table \ref{jointid}). Further, the reduction of one free parameter in case of LP with respect to BPL enforces the latter to be the most preferred particle distribution. In Table \ref{lp_par}, we give the best fit parameters for the case of the LP particle distribution. We see similar correlations as discussed in Section \ref{sect:photspec}. In fact, there is a strong linear correlation ($p< 1e-16$) between the corresponding parameters of the photon and the particle spectrum.

%
%
%

 \section[sect:Disc]{Discussions and Conclusions}\label{sect:Disc}

In the present work, we performed a detailed study of the bright X-ray flare of Mkn 421 observed during April 2013 along with information available at other wavebands. We noticed the X-ray flare is not significantly correlated with the UV and, in addition, the variability amplitude of the former is considerably larger than the latter. This suggests that probably the X-ray and UV emission may belong to emission from different particle distributions.

A detailed spectral analysis of the X-ray observations over different time periods during the flare suggests the emission to arise due to synchrotron mechanism from log parabola particle distribution. Though this particle distribution can be statistically more appealing, a broken power law particle spectrum cannot be excluded. Further, we extended the best fit LP and BPL particle distributions to low energies and predicted the UV synchrotron flux. During low X-ray flux states, the predicted UV flux agrees reasonably well with the observed flux. However, at high X-ray flux states, the observed UV flux is significantly higher that the predicted one by a factor of $ 2-50$, with the larger deviations corresponding to the LP model. In addition, the variability of the predicted UV flux is much higher ($F_{var,pred}=1.10 \pm 0.10$) than the one obtained from the observed UV flux. This study again questions the similar origin of X-ray and UV emission.

\begin{figure*}
	\centering
	\includegraphics[scale=0.9]{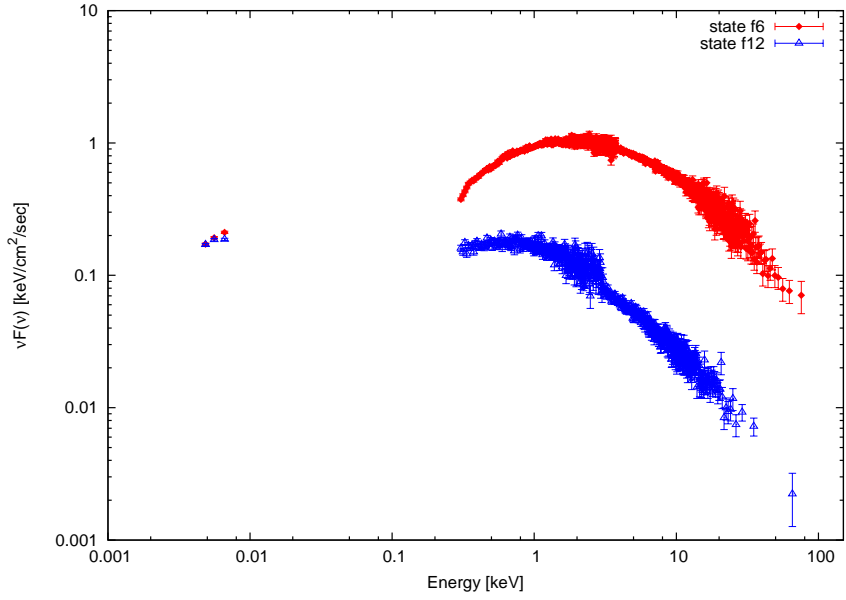}
	\caption{Observed X-ray and UV spectrum during two states. While the X-ray flux changes by more than a factor of 10, the UV flux remains constant.} 
	\label{flare_qu}
\end{figure*}

A plausible interpretation of this inconsistency between X-ray and UV fluxes can be done by associating the UV emission from the putative accretion disk. However, such thermal emission from the disk was never a important contribution in the UV bands for Mkn 421 \citep{fermi421}, and the UV spectral detail is not sufficient enough to assert this interpretation (Figure \ref{flare_qu}). Alternatively, the underlying particle distribution can be more complex than the ones studied in this work. Nevertheless, such a particle distribution demands a concave spectrum which is not possible with our present understanding of particle acceleration \citep{sunder_m87}. Hence, we attribute this unusual X-ray - UV behaviour of the source as a result of two population electron distribution. A similar conclusion was obtained by \cite{Mkn421_march2010} by studying a flare of the same source in March 2010. Such different electron distributions can be obtained if multiple emission regions are involved the emission process. If the flaring region is located at the recollimation zone of the jet, then a compact emission region can be achieved where the recollimation shock meets the jet axis \citep{tavecchio1222,Pankaj1222}. Alternatively, episodic particle acceleration suggested by \cite{perlman06} can be a reason for the second particle distribution. \cite{perlman} have shown that this can also explain the relative less variability of the optical/UV bands as compared to the X-ray.

\cite{Pian421} studied the same flare, starting from MJD 56397, with emphasis on {\it INTEGRAL} and {\it Fermi}-LAT data. They also found trends of spectral hardening with flux. However, unlike our results, the {\it INTEGRAL} spectral data could be well fit by a broken power law spectral form. They modelled the broadband SED with a simple one zone SSC model, and required large variations in the magnetic fields and Doppler factors to fit the SED of different states successfully. Also, our results do not match with that of \cite{massaro_lp1} as we do not see any correlation between $\alpha_s$ and $\beta_s$. Thus, our results are inconsistent with statistical particle acceleration.

\begin{figure*}
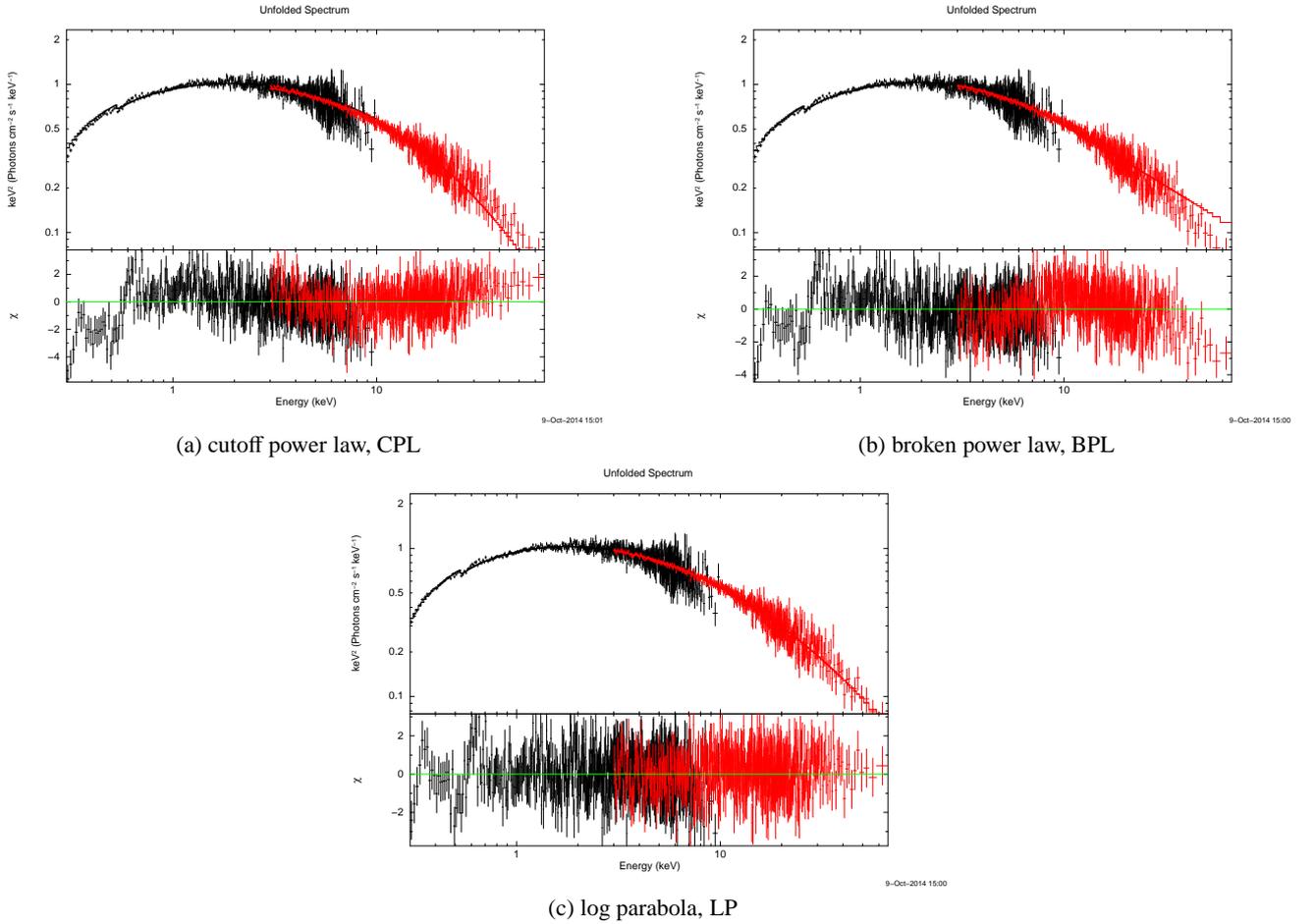

	\centering
	\subfloat[cutoff power law, CPL]{\includegraphics[scale=0.3, angle=270]{cutoffpl_eeufspec_delchi.eps}}
	\qquad
	\subfloat[broken power law, BPL]{\includegraphics[scale=0.3, angle=270]{bknpo_eeufspec_delchi.eps}}
	\qquad
	\subfloat[log parabola, LP]{\includegraphics[scale=0.3, angle=270]{logpar_eeufspec_delchi.eps}}
	\caption{Fitted spectrum and residuals for the three models for state {\bf f6}. The differences start showing up above $50keV$}
	\label{eeufchi19}
\end{figure*}

Our results indicate the potential of using broadband X-ray data to constrain the underlying particle spectrum and differentiate separate variable spectral components, especially if there are simultaneous data in other wavebands. Figure \ref{eeufchi19} shows the energy spectrum (in $\nu F_{\nu}$) and the residuals for the BPL, the CPL and the LP for the state {\bf f6}.  While the residuals show structure as a function of energy for energies $< 0.7$keV, especially for the CPL and the BPL, the data and model agree within $10\%$ at these energies. Moreover, we also implemented a fit at energies  $< 0.7$keV and verified that our results do not change significantly. Thus, this effect is unlikely to be a serious source of error in this work. More importantly, there is a clear systematic deviation from the data for the BPL and the  CPL at the higher energies, beyond $50keV$. The forthcoming satellite {\tt ASTROSAT}  \citep{astrosat}, with its wide band X-ray coverage and simultaneous optical-UV measurement can be expected to make significant breakthroughs in this field.

\begin{acknowledgements}
A. Sinha would like to thank Dr. Sunder Sahayanathan from the Bhabha Atomic Research Center, Mumbai, for helpful discussions and comments. This research has made use of data,
software and/or web tools obtained from NASAs High Energy Astrophysics Science Archive
Research Center (HEASARC), a service of Goddard Space Flight Center and the Smithsonian
Astrophysical Observatory. Part of this work is based on archival data, software, or online
services provided by the ASI Science Data Center (ASDC).	This research has made use of the
  XRT Data Analysis Software (XRTDAS) developed under the responsibility
the ASI Science Data Center (ASDC), Italy, and the  NuSTAR Data Analysis Software (NuSTARDAS) jointly developed by the ASI Science
    Data Center (ASDC, Italy) and the California Institute of Technology (Caltech, USA). 
	Data from the Steward Observatory spectropolarimetric monitoring project were used. This
program is supported by Fermi Guest Investigator grants NNX08AW56G, NNX09AU10G, and
NNX12AO93G.
\end{acknowledgements}

\bibliography{Mkn421}

\begin{thebibliography}{54}
\expandafter\ifx\csname natexlab\endcsname\relax\def\natexlab#1{#1}\fi

\bibitem[{{Abdo} {et~al.}(2011){Abdo}, {Ackermann}, {Ajello}, {Baldini},
  {Ballet}, {Barbiellini}, {Bastieri}, {Bechtol}, {Bellazzini}, {Berenji}, \&
  et~al.}]{fermi421}
{Abdo}, A.~A., {Ackermann}, M., {Ajello}, M., {et~al.} 2011, \apj, 736, 131

\bibitem[{{Acciari} {et~al.}(2009){Acciari}, {Aliu}, {Aune}, {Beilicke},
  {Benbow}, {B{\"o}ttcher}, {Bradbury}, {Buckley}, {Bugaev}, {Butt}, \&
  et~al.}]{MAGIC421_outburst}
{Acciari}, V.~A., {Aliu}, E., {Aune}, T., {et~al.} 2009, \apj, 703, 169

\bibitem[{{Albert} {et~al.}(2007){Albert}, {Aliu}, {Anderhub}, {Antoranz},
  {Armada}, {Asensio}, {Baixeras}, {Barrio}, {Bartko}, {Bastieri}, {Becker},
  {Bednarek}, {Berger}, {Bigongiari}, {Biland}, {Bock}, {Bordas},
  {Bosch-Ramon}, {Bretz}, {Britvitch}, {Camara}, {Carmona}, {Chilingarian},
  {Ciprini}, {Coarasa}, {Commichau}, {Contreras}, {Cortina}, {Curtef},
  {Danielyan}, {Dazzi}, {De Angelis}, {de los Reyes}, {De Lotto},
  {Domingo-Santamar{\'{\i}}a}, {Dorner}, {Doro}, {Errando}, {Fagiolini},
  {Ferenc}, {Fern{\'a}ndez}, {Firpo}, {Flix}, {Fonseca}, {Font}, {Fuchs},
  {Galante}, {Garczarczyk}, {Gaug}, {Giller}, {Goebel}, {Hakobyan},
  {Hayashida}, {Hengstebeck}, {H{\"o}hne}, {Hose}, {Hsu}, {Jacon}, {Jogler},
  {Kalekin}, {Kosyra}, {Kranich}, {Kritzer}, {Laatiaoui}, {Laille}, {Liebing},
  {Lindfors}, {Lombardi}, {Longo}, {L{\'o}pez}, {L{\'o}pez}, {Lorenz},
  {Majumdar}, {Maneva}, {Mannheim}, {Mansutti}, {Mariotti}, {Mart{\'{\i}}nez},
  {Mazin}, {Merck}, {Meucci}, {Meyer}, {Miranda}, {Mirzoyan}, {Mizobuchi},
  {Moralejo}, {Nilsson}, {Ninkovic}, {O{\~n}a-Wilhelmi}, {Ordu{\~n}a}, {Otte},
  {Oya}, {Paneque}, {Paoletti}, {Paredes}, {Pasanen}, {Pascoli}, {Pauss},
  {Pegna}, {Persic}, {Peruzzo}, {Piccioli}, {Poller}, {Prandini}, {Raymers},
  {Rhode}, {Rib{\'o}}, {Rico}, {Rissi}, {Robert}, {R{\"u}gamer}, {Saggion},
  {S{\'a}nchez}, {Sartori}, {Scalzotto}, {Scapin}, {Schmitt}, {Schweizer},
  {Shayduk}, {Shinozaki}, {Shore}, {Sidro}, {Sillanp{\"a}{\"a}}, {Sobczynska},
  {Stamerra}, {Stark}, {Takalo}, {Temnikov}, {Tescaro}, {Teshima}, {Tonello},
  {Torres}, {Torres}, {Turini}, {Vankov}, {Vitale}, {Wagner}, {Wibig},
  {Wittek}, {Zanin}, \& {Zapatero}}]{MAGIC421}
{Albert}, J., {Aliu}, E., {Anderhub}, H., {et~al.} 2007, \apj, 663, 125

\bibitem[{{Aleksi{\'c}} {et~al.}(2012){Aleksi{\'c}}, {Alvarez}, {Antonelli},
  {Antoranz}, {Asensio}, {Backes}, {Barrio}, {Bastieri}, {Becerra
  Gonz{\'a}lez}, {Bednarek}, {Berdyugin}, {Berger}, {Bernardini}, {Biland},
  {Blanch}, {Bock}, {Boller}, {Bonnoli}, {Borla Tridon}, {Braun}, {Bretz},
  {Ca{\~n}ellas}, {Carmona}, {Carosi}, {Colin}, {Colombo}, {Contreras},
  {Cortina}, {Cossio}, {Covino}, {Dazzi}, {De Angelis}, {De Caneva}, {De Cea
  del Pozo}, {De Lotto}, {Delgado Mendez}, {Diago Ortega}, {Doert},
  {Dom{\'{\i}}nguez}, {Dominis Prester}, {Dorner}, {Doro}, {Elsaesser},
  {Ferenc}, {Fonseca}, {Font}, {Fruck}, {Garc{\'{\i}}a L{\'o}pez},
  {Garczarczyk}, {Garrido}, {Giavitto}, {Godinovi{\'c}}, {Hadasch},
  {H{\"a}fner}, {Herrero}, {Hildebrand}, {H{\"o}hne-M{\"o}nch}, {Hose},
  {Hrupec}, {Huber}, {Jogler}, {Kellermann}, {Klepser}, {Kr{\"a}henb{\"u}hl},
  {Krause}, {La Barbera}, {Lelas}, {Leonardo}, {Lindfors}, {Lombardi},
  {L{\'o}pez}, {L{\'o}pez}, {Lorenz}, {Makariev}, {Maneva}, {Mankuzhiyil},
  {Mannheim}, {Maraschi}, {Mariotti}, {Mart{\'{\i}}nez}, {Mazin}, {Meucci},
  {Miranda}, {Mirzoyan}, {Miyamoto}, {Mold{\'o}n}, {Moralejo}, {Munar-Adrover},
  {Nieto}, {Nilsson}, {Orito}, {Oya}, {Paneque}, {Paoletti}, {Pardo},
  {Paredes}, {Partini}, {Pasanen}, {Pauss}, {Perez-Torres}, {Persic},
  {Peruzzo}, {Pilia}, {Pochon}, {Prada}, {Prada Moroni}, {Prandini}, {Puljak},
  {Reichardt}, {Reinthal}, {Rhode}, {Rib{\'o}}, {Rico}, {R{\"u}gamer},
  {Saggion}, {Saito}, {Saito}, {Salvati}, {Satalecka}, {Scalzotto}, {Scapin},
  {Schultz}, {Schweizer}, {Shayduk}, {Shore}, {Sillanp{\"a}{\"a}}, {Sitarek},
  {Sobczynska}, {Spanier}, {Spiro}, {Stamerra}, {Steinke}, {Storz}, {Strah},
  {Suri{\'c}}, {Takalo}, {Takami}, {Tavecchio}, {Temnikov}, {Terzi{\'c}},
  {Tescaro}, {Teshima}, {Tibolla}, {Torres}, {Treves}, {Uellenbeck}, {Vankov},
  {Vogler}, {Wagner}, {Weitzel}, {Zabalza}, {Zandanel}, \&
  {Zanin}}]{MAGIC421_2008}
{Aleksi{\'c}}, J., {Alvarez}, E.~A., {Antonelli}, L.~A., {et~al.} 2012, \aap,
  542, A100

\bibitem[{{Aleksi{\'c}} {et~al.}(2010){Aleksi{\'c}}, {Anderhub}, {Antonelli},
  {Antoranz}, {Backes}, {Baixeras}, {Balestra}, {Barrio}, {Bastieri}, {Becerra
  Gonz{\'a}lez}, {Becker}, {Bednarek}, {Berdyugin}, {Berger}, {Bernardini},
  {Biland}, {Bock}, {Bonnoli}, {Bordas}, {Borla Tridon}, {Bosch-Ramon}, {Bose},
  {Braun}, {Bretz}, {Britzger}, {Camara}, {Carmona}, {Carosi}, {Colin},
  {Commichau}, {Contreras}, {Cortina}, {Costado}, {Covino}, {Dazzi}, {de
  Angelis}, {de Cea Del Pozo}, {de Los Reyes}, {de Lotto}, {de Maria}, {de
  Sabata}, {Delgado Mendez}, {Doert}, {Dom{\'{\i}}nguez}, {Dominis Prester},
  {Dorner}, {Doro}, {Elsaesser}, {Errando}, {Ferenc}, {Fonseca}, {Font},
  {Garc{\'{\i}}a L{\'o}pez}, {Garczarczyk}, {Gaug}, {Godinovic}, {Hadasch},
  {Herrero}, {Hildebrand}, {H{\"o}hne-M{\"o}nch}, {Hose}, {Hrupec}, {Hsu},
  {Jogler}, {Klepser}, {Kr{\"a}henb{\"u}hl}, {Kranich}, {La Barbera}, {Laille},
  {Leonardo}, {Lindfors}, {Lombardi}, {Longo}, {L{\'o}pez}, {Lorenz},
  {Majumdar}, {Maneva}, {Mankuzhiyil}, {Mannheim}, {Maraschi}, {Mariotti},
  {Mart{\'{\i}}nez}, {Mazin}, {Meucci}, {Miranda}, {Mirzoyan}, {Miyamoto},
  {Mold{\'o}n}, {Moles}, {Moralejo}, {Nieto}, {Nilsson}, {Ninkovic}, {Orito},
  {Oya}, {Paoletti}, {Paredes}, {Partini}, {Pasanen}, {Pascoli}, {Pauss},
  {Pegna}, {Perez-Torres}, {Persic}, {Peruzzo}, {Prada}, {Prandini},
  {Puchades}, {Puljak}, {Reichardt}, {Rhode}, {Rib{\'o}}, {Rico}, {Rissi},
  {R{\"u}gamer}, {Saggion}, {Saito}, {Salvati}, {S{\'a}nchez-Conde},
  {Satalecka}, {Scalzotto}, {Scapin}, {Schweizer}, {Shayduk}, {Shore},
  {Sierpowska-Bartosik}, {Sillanp{\"a}{\"a}}, {Sitarek}, {Sobczynska},
  {Spanier}, {Spiro}, {Stamerra}, {Steinke}, {Strah}, {Struebig}, {Suric},
  {Takalo}, {Tavecchio}, {Temnikov}, {Tescaro}, {Teshima}, {Torres}, {Vankov},
  {Wagner}, {Zabalza}, {Zandanel}, {Zanin}, \& {MAGIC
  Collaboration}}]{MAGIC421_2006}
{Aleksi{\'c}}, J., {Anderhub}, H., {Antonelli}, L.~A., {et~al.} 2010, \aap,
  519, A32

\bibitem[{{Aleksi{\'c}} {et~al.}(2014){Aleksi{\'c}}, {Ansoldi}, {Antonelli},
  {Antoranz}, {Babic}, {Bangale}, {Barres de Almeida}, {Barrio}, {Becerra
  Gonz{\'a}lez}, \& et~al.}]{Mkn421_march2010}
{Aleksi{\'c}}, J., {Ansoldi}, S., {Antonelli}, L.~A., {et~al.} 2014, ArXiv
  e-prints

\bibitem[{{Alexander}(1997)}]{zdcf}
{Alexander}, T. 1997, in Astrophysics and Space Science Library, Vol. 218,
  Astronomical Time Series, ed. D.~{Maoz}, A.~{Sternberg}, \& E.~M.
  {Leibowitz}, 163

\bibitem[{{Balokovi{\'c}} {et~al.}(2013){Balokovi{\'c}}, {Ajello}, {Blandford},
  {Boggs}, {Borracci}, {Chiang}, {Christensen}, {Craig}, {Forster}, {Furniss},
  {F{\"u}rst}, {Ghisellini}, {Giebels}, {Giommi}, {Grefenstette}, {Hailey},
  {Harrison}, {Hayashida}, {Humensky}, {Inoue}, {Koglin}, {Krawczynski},
  {Madejski}, {Madsen}, {Meier}, {Nelson}, {Ogle}, {Paneque}, {Perri},
  {Puccetti}, {Reynolds}, {Sbarrato}, {Stern}, {Tagliaferri}, {Urry}, {Wehrle},
  \& {Zhang}}]{nustar421}
{Balokovi{\'c}}, M., {Ajello}, M., {Blandford}, R.~D., {et~al.} 2013, in
  European Physical Journal Web of Conferences, Vol.~61, European Physical
  Journal Web of Conferences, 4013

\bibitem[{{B{\l}a{\.z}ejowski} {et~al.}(2005){B{\l}a{\.z}ejowski}, {Blaylock},
  {Bond}, {Bradbury}, {Buckley}, {Carter-Lewis}, {Celik}, {Cogan}, {Cui},
  {Daniel}, {Duke}, {Falcone}, {Fegan}, {Fegan}, {Finley}, {Fortson},
  {Gammell}, {Gibbs}, {Gillanders}, {Grube}, {Gutierrez}, {Hall}, {Hanna},
  {Holder}, {Horan}, {Humensky}, {Kenny}, {Kertzman}, {Kieda}, {Kildea},
  {Knapp}, {Kosack}, {Krawczynski}, {Krennrich}, {Lang}, {LeBohec}, {Linton},
  {Lloyd-Evans}, {Maier}, {Mendoza}, {Milovanovic}, {Moriarty}, {Nagai}, {Ong},
  {Power-Mooney}, {Quinn}, {Quinn}, {Ragan}, {Reynolds}, {Rebillot}, {Rose},
  {Schroedter}, {Sembroski}, {Swordy}, {Syson}, {Valcarel}, {Vassiliev},
  {Wakely}, {Walker}, {Weekes}, {White}, {Zweerink}, {VERITAS Collaboration},
  {Mochejska}, {Smith}, {Aller}, {Aller}, {Ter{\"a}sranta}, {Boltwood},
  {Sadun}, {Stanek}, {Adams}, {Foster}, {Hartman}, {Lai}, {B{\"o}ttcher},
  {Reimer}, \& {Jung}}]{multi421}
{B{\l}a{\.z}ejowski}, M., {Blaylock}, G., {Bond}, I.~H., {et~al.} 2005, \apj,
  630, 130

\bibitem[{{B{\"o}ttcher}(2007)}]{bottcher2007}
{B{\"o}ttcher}, M. 2007, \apss, 307, 69

\bibitem[{{Breeveld} {et~al.}(2011){Breeveld}, {Landsman}, {Holland}, {Roming},
  {Kuin}, \& {Page}}]{UVOT_conv}
{Breeveld}, A.~A., {Landsman}, W., {Holland}, S.~T., {et~al.} 2011, in American
  Institute of Physics Conference Series, Vol. 1358, American Institute of
  Physics Conference Series, ed. J.~E. {McEnery}, J.~L. {Racusin}, \&
  N.~{Gehrels}, 373--376

\bibitem[{{Brinkmann} {et~al.}(2003){Brinkmann}, {Papadakis}, {den Herder}, \&
  {Haberl}}]{XMM421_2003}
{Brinkmann}, W., {Papadakis}, I.~E., {den Herder}, J.~W.~A., \& {Haberl}, F.
  2003, \aap, 402, 929

\bibitem[{{Burrows} {et~al.}(2005){Burrows}, {Hill}, {Nousek}, {Kennea},
  {Wells}, {Osborne}, {Abbey}, {Beardmore}, {Mukerjee}, {Short}, {Chincarini},
  {Campana}, {Citterio}, {Moretti}, {Pagani}, {Tagliaferri}, {Giommi},
  {Capalbi}, {Tamburelli}, {Angelini}, {Cusumano}, {Br{\"a}uninger}, {Burkert},
  \& {Hartner}}]{XRT_instru}
{Burrows}, D.~N., {Hill}, J.~E., {Nousek}, J.~A., {et~al.} 2005, \ssr, 120, 165

\bibitem[{{Chitnis} {et~al.}(2009){Chitnis}, {Pendharkar}, {Bose}, {Agrawal},
  {Rao}, \& {Misra}}]{varsha_var}
{Chitnis}, V.~R., {Pendharkar}, J.~K., {Bose}, D., {et~al.} 2009, \apj, 698,
  1207

\bibitem[{{Foschini} {et~al.}(2011){Foschini}, {Ghisellini}, {Tavecchio},
  {Bonnoli}, \& {Stamerra}}]{fluxdoubling}
{Foschini}, L., {Ghisellini}, G., {Tavecchio}, F., {Bonnoli}, G., \&
  {Stamerra}, A. 2011, \aap, 530, A77

\bibitem[{{Fossati} {et~al.}(2008){Fossati}, {Buckley}, {Bond}, {Bradbury},
  {Carter-Lewis}, {Chow}, {Cui}, {Falcone}, {Finley}, {Gaidos}, {Grube},
  {Holder}, {Horan}, {Horns}, {Jordan}, {Kieda}, {Kildea}, {Krawczynski},
  {Krennrich}, {Lang}, {LeBohec}, {Lee}, {Moriarty}, {Ong}, {Petry}, {Quinn},
  {Sembroski}, {Wakely}, \& {Weekes}}]{421_march2001}
{Fossati}, G., {Buckley}, J.~H., {Bond}, I.~H., {et~al.} 2008, \apj, 677, 906

\bibitem[{{Fossati} {et~al.}(2000{\natexlab{a}}){Fossati}, {Celotti},
  {Chiaberge}, {Zhang}, {Chiappetti}, {Ghisellini}, {Maraschi}, {Tavecchio},
  {Pian}, \& {Treves}}]{BeppoSAX-fossati}
{Fossati}, G., {Celotti}, A., {Chiaberge}, M., {et~al.} 2000{\natexlab{a}},
  \apj, 541, 153

\bibitem[{{Fossati} {et~al.}(2000{\natexlab{b}}){Fossati}, {Celotti},
  {Chiaberge}, {Zhang}, {Chiappetti}, {Ghisellini}, {Maraschi}, {Tavecchio},
  {Pian}, \& {Treves}}]{fossati421}
{Fossati}, G., {Celotti}, A., {Chiaberge}, M., {et~al.} 2000{\natexlab{b}},
  \apj, 541, 166

\bibitem[{{Harrison} {et~al.}(2013){Harrison}, {Craig}, {Christensen},
  {Hailey}, {Zhang}, {Boggs}, {Stern}, {Cook}, {Forster}, {Giommi},
  {Grefenstette}, {Kim}, {Kitaguchi}, {Koglin}, {Madsen}, {Mao}, {Miyasaka},
  {Mori}, {Perri}, {Pivovaroff}, {Puccetti}, {Rana}, {Westergaard}, {Willis},
  {Zoglauer}, {An}, {Bachetti}, {Barri{\`e}re}, {Bellm}, {Bhalerao},
  {Brejnholt}, {Fuerst}, {Liebe}, {Markwardt}, {Nynka}, {Vogel}, {Walton},
  {Wik}, {Alexander}, {Cominsky}, {Hornschemeier}, {Hornstrup}, {Kaspi},
  {Madejski}, {Matt}, {Molendi}, {Smith}, {Tomsick}, {Ajello}, {Ballantyne},
  {Balokovi{\'c}}, {Barret}, {Bauer}, {Blandford}, {Brandt}, {Brenneman},
  {Chiang}, {Chakrabarty}, {Chenevez}, {Comastri}, {Dufour}, {Elvis}, {Fabian},
  {Farrah}, {Fryer}, {Gotthelf}, {Grindlay}, {Helfand}, {Krivonos}, {Meier},
  {Miller}, {Natalucci}, {Ogle}, {Ofek}, {Ptak}, {Reynolds}, {Rigby},
  {Tagliaferri}, {Thorsett}, {Treister}, \& {Urry}}]{Nustar_instru}
{Harrison}, F.~A., {Craig}, W.~W., {Christensen}, F.~E., {et~al.} 2013, \apj,
  770, 103

\bibitem[{{Horan} {et~al.}(2009){Horan}, {Acciari}, {Bradbury}, {Buckley},
  {Bugaev}, {Byrum}, {Cannon}, {Celik}, {Cesarini}, {Chow}, {Ciupik}, {Cogan},
  {Falcone}, {Fegan}, {Finley}, {Fortin}, {Fortson}, {Gall}, {Gillanders},
  {Grube}, {Gyuk}, {Hanna}, {Hays}, {Kertzman}, {Kildea}, {Konopelko},
  {Krawczynski}, {Krennrich}, {Lang}, {Lee}, {Moriarty}, {Nagai}, {Niemiec},
  {Ong}, {Perkins}, {Pohl}, {Quinn}, {Reynolds}, {Rose}, {Sembroski}, {Smith},
  {Steele}, {Swordy}, {Toner}, {Vassiliev}, {Wakely}, {Weekes}, {White},
  {Williams}, {Wood}, {Zitzer}, {Aller}, {Aller}, {Baker}, {Barnaby}, {Carini},
  {Charlot}, {Dumm}, {Fields}, {Hovatta}, {Jordan}, {Kovalev}, {Kovalev},
  {Krimm}, {Kurtanidze}, {L{\"a}hteenm{\"a}ki}, {LeCampion}, {Maune},
  {Montaruli}, {Sadun}, {Smith}, {Tornikoski}, {Turunen}, \&
  {Walters}}]{multi421_2005-2006}
{Horan}, D., {Acciari}, V.~A., {Bradbury}, S.~M., {et~al.} 2009, \apj, 695, 596

\bibitem[{{Isobe} {et~al.}(2010){Isobe}, {Sugimori}, {Kawai}, {Ueda}, {Negoro},
  {Sugizaki}, {Matsuoka}, {Daikyuji}, {Eguchi}, {Hiroi}, {Ishikawa},
  {Ishiwata}, {Kawasaki}, {Kimura}, {Kohama}, {Mihara}, {Miyoshi}, {Morii},
  {Nakagawa}, {Nakahira}, {Nakajima}, {Ozawa}, {Sootome}, {Suzuki}, {Tomida},
  {Tsunemi}, {Ueno}, {Yamamoto}, {Yamaoka}, {Yoshida}, \& {MAXI
  Team}}]{MAXI421}
{Isobe}, N., {Sugimori}, K., {Kawai}, N., {et~al.} 2010, \pasj, 62, L55

\bibitem[{jie Qian {et~al.}(1998)jie Qian, zhen Zhang, Witzel, Krichbaum,
  Britzen, \& Kraus}]{Qian}
jie Qian, S., zhen Zhang, X., Witzel, A., {et~al.} 1998, Chinese Astronomy and
  Astrophysics, 22, 155

\bibitem[{{K. Katarzynski} {et~al.}(2005){K. Katarzynski}, {G. Ghisellini}, {F.
  Tavecchio}, {L. Maraschi}, {G. Fossati}, \& {A.
  Mastichiadis}}]{Katarzynski_hbl}
{K. Katarzynski}, {G. Ghisellini}, {F. Tavecchio}, {et~al.} 2005, A\&A, 433,
  479

\bibitem[{{Kalberla} {et~al.}(2005){Kalberla}, {Burton}, {Hartmann}, {Arnal},
  {Bajaja}, {Morras}, \& {P{\"o}ppel}}]{LAB}
{Kalberla}, P.~M.~W., {Burton}, W.~B., {Hartmann}, D., {et~al.} 2005, \aap,
  440, 775

\bibitem[{{Krawczynski} {et~al.}(2001){Krawczynski}, {Sambruna}, {Kohnle},
  {Coppi}, {Aharonian}, {Akhperjanian}, {Barrio}, {Bernl{\"o}hr}, {B{\"o}rst},
  {Bojahr}, {Bolz}, {Contreras}, {Cortina}, {Denninghoff}, {Fonseca},
  {Gonzalez}, {G{\"o}tting}, {Heinzelmann}, {Hermann}, {Heusler}, {Hofmann},
  {Horns}, {Ibarra}, {Jung}, {Kankanyan}, {Kestel}, {Kettler}, {Konopelko},
  {Kornmeyer}, {Kranich}, {Lampeitl}, {Lorenz}, {Lucarelli}, {Magnussen},
  {Mang}, {Meyer}, {Mirzoyan}, {Moralejo}, {Padilla}, {Panter}, {Plaga},
  {Plyasheshnikov}, {P{\"u}hlhofer}, {Rauterberg}, {R{\"o}hring}, {Rhode},
  {Rowell}, {Sahakian}, {Samorski}, {Schilling}, {Schr{\"o}der}, {Siems},
  {Stamm}, {Tluczykont}, {V{\"o}lk}, {Wiedner}, \& {Wittek}}]{Mkn421_2000}
{Krawczynski}, H., {Sambruna}, R., {Kohnle}, A., {et~al.} 2001, \apj, 559, 187

\bibitem[{{Kushwaha} {et~al.}(2014){Kushwaha}, {Sahayanathan}, {Lekshmi},
  {Singh}, {Bhattacharyya}, \& {Bhattacharya}}]{Pankaj1222}
{Kushwaha}, P., {Sahayanathan}, S., {Lekshmi}, R., {et~al.} 2014, \mnras, 442,
  131

\bibitem[{{Li} {et~al.}(2013){Li}, {Zhang}, {Zhang}, {Xiong}, {Liu}, {Cha}, \&
  {Li}}]{Xray-GeV_corr}
{Li}, B., {Zhang}, H., {Zhang}, X., {et~al.} 2013, \apss, 347, 349

\bibitem[{{Lichti} {et~al.}(2008){Lichti}, {Bottacini}, {Ajello}, {Charlot},
  {Collmar}, {Falcone}, {Horan}, {Huber}, {von Kienlin}, {L{\"a}hteenm{\"a}ki},
  {Lindfors}, {Morris}, {Nilsson}, {Petry}, {R{\"u}ger}, {Sillanp{\"a}{\"a}},
  {Spanier}, \& {Tornikoski}}]{Integral421}
{Lichti}, G.~G., {Bottacini}, E., {Ajello}, M., {et~al.} 2008, \aap, 486, 721

\bibitem[{{Massaro} {et~al.}(2004){Massaro}, {Perri}, {Giommi}, \&
  {Nesci}}]{massaro_lp1}
{Massaro}, E., {Perri}, M., {Giommi}, P., \& {Nesci}, R. 2004, \aap, 413, 489

\bibitem[{{Massaro} {et~al.}(2008){Massaro}, {Tramacere}, {Cavaliere}, {Perri},
  \& {Giommi}}]{lp_many}
{Massaro}, F., {Tramacere}, A., {Cavaliere}, A., {Perri}, M., \& {Giommi}, P.
  2008, \aap, 478, 395

\bibitem[{{Nolan} {et~al.}(2012){Nolan}, {Abdo}, {Ackermann}, {Ajello},
  {Allafort}, {Antolini}, {Atwood}, {Axelsson}, {Baldini}, {Ballet}, \&
  et~al.}]{2FGL}
{Nolan}, P.~L., {Abdo}, A.~A., {Ackermann}, M., {et~al.} 2012, \apjs, 199, 31

\bibitem[{{Padovani} \& {Giommi}(1995)}]{PadGio}
{Padovani}, P. \& {Giommi}, P. 1995, \apj, 444, 567

\bibitem[{{Paliya} {et~al.}(2015){Paliya}, {Sahayanathan}, \&
  {Stalin}}]{Vaidehi3C279}
{Paliya}, V.~S., {Sahayanathan}, S., \& {Stalin}, C.~S. 2015, ArXiv e-prints

\bibitem[{{Perlman} {et~al.}(2006){Perlman}, {Daugherty}, {Georganopoulos},
  {Koratkar}, {Madejski}, {Andersson}, {Krolik}, {Rector}, {Stocke}, {Wagner},
  {Aller}, {Aller}, \& {Allen}}]{perlman06}
{Perlman}, E.~S., {Daugherty}, T., {Georganopoulos}, M., {et~al.} 2006, in
  Astronomical Society of the Pacific Conference Series, Vol. 350, Blazar
  Variability Workshop II: Entering the GLAST Era, ed. H.~R. {Miller},
  K.~{Marshall}, J.~R. {Webb}, \& M.~F. {Aller}, 191

\bibitem[{{Perlman} {et~al.}(2005){Perlman}, {Madejski}, {Georganopoulos},
  {Andersson}, {Daugherty}, {Krolik}, {Rector}, {Stocke}, {Koratkar}, {Wagner},
  {Aller}, {Aller}, \& {Allen}}]{perlman}
{Perlman}, E.~S., {Madejski}, G., {Georganopoulos}, M., {et~al.} 2005, \apj,
  625, 727

\bibitem[{{Pian} {et~al.}(2014){Pian}, {T{\"u}rler}, {Fiocchi}, {Boissay},
  {Bazzano}, {Foschini}, {Tavecchio}, {Bianchin}, {Castignani}, {Ferrigno},
  {Raiteri}, {Villata}, {Beckmann}, {D'Ammando}, {Hudec}, {Malaguti},
  {Maraschi}, {Pursimo}, {Romano}, {Soldi}, {Stamerra}, {Treves}, {Ubertini},
  {Vercellone}, \& {Walter}}]{Pian421}
{Pian}, E., {T{\"u}rler}, M., {Fiocchi}, M., {et~al.} 2014, \aap, 570, A77

\bibitem[{{Punch} {et~al.}(1992){Punch}, {Akerlof}, {Cawley}, {Chantell},
  {Fegan}, {Fennell}, {Gaidos}, {Hagan}, {Hillas}, {Jiang}, {Kerrick}, {Lamb},
  {Lawrence}, {Lewis}, {Meyer}, {Mohanty}, {O'Flaherty}, {Reynolds}, {Rovero},
  {Schubnell}, {Sembroski}, {Weekes}, \& {Wilson}}]{punch}
{Punch}, M., {Akerlof}, C.~W., {Cawley}, M.~F., {et~al.} 1992, \nat, 358, 477

\bibitem[{{Romano} {et~al.}(2006){Romano}, {Campana}, {Chincarini}, {Cummings},
  {Cusumano}, {Holland}, {Mangano}, {Mineo}, {Page}, {Pal'Shin}, {Rol},
  {Sakamoto}, {Zhang}, {Aptekar}, {Barbier}, {Barthelmy}, {Beardmore}, {Boyd},
  {Burrows}, {Capalbi}, {Fenimore}, {Frederiks}, {Gehrels}, {Giommi}, {Goad},
  {Godet}, {Golenetskii}, {Guetta}, {Kennea}, {La Parola}, {Malesani},
  {Marshall}, {Moretti}, {Nousek}, {O'Brien}, {Osborne}, {Perri}, \&
  {Tagliaferri}}]{XRT_pileup}
{Romano}, P., {Campana}, S., {Chincarini}, G., {et~al.} 2006, \aap, 456, 917

\bibitem[{{Roming} {et~al.}(2005){Roming}, {Kennedy}, {Mason}, {Nousek}, {Ahr},
  {Bingham}, {Broos}, {Carter}, {Hancock}, {Huckle}, {Hunsberger}, {Kawakami},
  {Killough}, {Koch}, {McLelland}, {Smith}, {Smith}, {Soto}, {Boyd},
  {Breeveld}, {Holland}, {Ivanushkina}, {Pryzby}, {Still}, \&
  {Stock}}]{UVOT_instru}
{Roming}, P.~W.~A., {Kennedy}, T.~E., {Mason}, K.~O., {et~al.} 2005, \ssr, 120,
  95

\bibitem[{{Sahayanathan}(2008)}]{sunder_m87}
{Sahayanathan}, S. 2008, \mnras, 388, L49

\bibitem[{{Schlegel} {et~al.}(1998){Schlegel}, {Finkbeiner}, \&
  {Davis}}]{schelgel}
{Schlegel}, D.~J., {Finkbeiner}, D.~P., \& {Davis}, M. 1998, \apj, 500, 525

\bibitem[{{Shukla} {et~al.}(2012){Shukla}, {Chitnis}, {Vishwanath}, {Acharya},
  {Anupama}, {Bhattacharjee}, {Britto}, {Prabhu}, {Saha}, \& {Singh}}]{Amit421}
{Shukla}, A., {Chitnis}, V.~R., {Vishwanath}, P.~R., {et~al.} 2012, \aap, 541,
  A140

\bibitem[{{Singh} {et~al.}(2014){Singh}, {Tandon}, {Agrawal}, {Antia},
  {Manchanda}, {Yadav}, {Seetha}, {Ramadevi}, {Rao}, {Bhattacharya}, {Paul},
  {Sreekumar}, {Bhattacharyya}, {Stewart}, {Hutchings}, {Annapurni}, {Ghosh},
  {Murthy}, {Pati}, {Rao}, {Stalin}, {Girish}, {Sankarasubramanian},
  {Vadawale}, {Bhalerao}, {Dewangan}, {Dedhia}, {Hingar}, {Katoch}, {Kothare},
  {Mirza}, {Mukerjee}, {Shah}, {Shah}, {Mohan}, {Sangal}, {Nagabhusana},
  {Sriram}, {Malkar}, {Sreekumar}, {Abbey}, {Hansford}, {Beardmore}, {Sharma},
  {Murthy}, {Kulkarni}, {Meena}, {Babu}, \& {Postma}}]{astrosat}
{Singh}, K.~P., {Tandon}, S.~N., {Agrawal}, P.~C., {et~al.} 2014, in Society of
  Photo-Optical Instrumentation Engineers (SPIE) Conference Series, Vol. 9144,
  Society of Photo-Optical Instrumentation Engineers (SPIE) Conference Series,
  1

\bibitem[{{Smith} {et~al.}(2009){Smith}, {Montiel}, {Rightley}, {Turner},
  {Schmidt}, \& {Jannuzi}}]{CCD-SPOL}
{Smith}, P.~S., {Montiel}, E., {Rightley}, S., {et~al.} 2009, ArXiv e-prints

\bibitem[{{Tavecchio} {et~al.}(2011){Tavecchio}, {Becerra-Gonzalez},
  {Ghisellini}, {Stamerra}, {Bonnoli}, {Foschini}, \&
  {Maraschi}}]{tavecchio1222}
{Tavecchio}, F., {Becerra-Gonzalez}, J., {Ghisellini}, G., {et~al.} 2011, \aap,
  534, A86

\bibitem[{{Tavecchio} {et~al.}(1998){Tavecchio}, {Maraschi}, \&
  {Ghisellini}}]{tavKN}
{Tavecchio}, F., {Maraschi}, L., \& {Ghisellini}, G. 1998, \apj, 509, 608

\bibitem[{Tomsick {et~al.}(2014)Tomsick, Nowak, Parker, Miller, Fabian,
  Harrison, Bachetti, Barret, Boggs, Christensen, Craig, Forster, Fürst,
  Grefenstette, Hailey, King, Madsen, Natalucci, Pottschmidt, Ross, Stern,
  Walton, Wilms, \& Zhang}]{harddefn}
Tomsick, J.~A., Nowak, M.~A., Parker, M., {et~al.} 2014, The Astrophysical
  Journal, 780, 78

\bibitem[{{Tramacere} {et~al.}(2007){Tramacere}, {Giommi}, {Massaro}, {Perri},
  {Nesci}, {Colafrancesco}, {Tagliaferri}, {Chincarini}, {Falcone}, {Burrows},
  {Roming}, {McMath Chester}, \& {Gehrels}}]{swift_lp}
{Tramacere}, A., {Giommi}, P., {Massaro}, E., {et~al.} 2007, \aap, 467, 501

\bibitem[{{Tramacere} {et~al.}(2009){Tramacere}, {Giommi}, {Perri},
  {Verrecchia}, \& {Tosti}}]{Swift421_2006}
{Tramacere}, A., {Giommi}, P., {Perri}, M., {Verrecchia}, F., \& {Tosti}, G.
  2009, \aap, 501, 879

\bibitem[{{Urry} \& {Padovani}(1995)}]{UrryPadovani}
{Urry}, C.~M. \& {Padovani}, P. 1995, \pasp, 107, 803

\bibitem[{{Ushio} {et~al.}(2009){Ushio}, {Tanaka}, {Madejski}, {Takahashi},
  {Hayashida}, {Kataoka}, {Mazin}, {R{\"u}gamer}, {Sato}, {Teshima}, {Wagner},
  \& {Yaji}}]{Suzaku421_2006}
{Ushio}, M., {Tanaka}, T., {Madejski}, G., {et~al.} 2009, \apj, 699, 1964

\bibitem[{{Vaughan} {et~al.}(2003){Vaughan}, {Edelson}, {Warwick}, \&
  {Uttley}}]{Vaughan}
{Vaughan}, S., {Edelson}, R., {Warwick}, R.~S., \& {Uttley}, P. 2003, \mnras,
  345, 1271

\bibitem[{{W. Brinkmann} {et~al.}(2001){W. Brinkmann}, {S. Sembay}, {R. G.
  Griffiths}, {G. Branduardi-Raymont}, {M. Gliozzi}, {Th. Boller}, {A. Tiengo},
  {S. Molendi}, \& {S. Zane}}]{XMM421}
{W. Brinkmann}, {S. Sembay}, {R. G. Griffiths}, {et~al.} 2001, A\&A, 365, L162

\bibitem[{{Zhang} {et~al.}(2005){Zhang}, {Treves}, {Celotti}, {Qin}, \&
  {Bai}}]{Zhang2155}
{Zhang}, Y.~H., {Treves}, A., {Celotti}, A., {Qin}, Y.~P., \& {Bai}, J.~M.
  2005, \apj, 629, 686

\end{thebibliography}
\bibliographystyle{aa}


\begin{table*}
	\centering
	\caption{Details of Nustar pointings}
	\begin{tabular}{  c    c    c     }
		\hline
	    \hline

		obsid        &  start date and time &  exposure (in sec)  \\ \hline 
		60002023023  &  2013-04-10   20:53:07  &      118  \\ 
		60002023024  &  2013-04-10   21:26:07  &     5758  \\ 
		60002023025  &  2013-04-11   01:01:07  &    57509  \\ 
		60002023026  &  2013-04-12   20:11:07  &      441  \\ 
		60002023027  &  2013-04-12   20:36:07  &     7630  \\ 
		60002023029  &  2013-04-13   21:36:07  &    16510  \\ 
		60002023031  &  2013-04-14   21:41:07  &    15606  \\ 
		60002023033  &  2013-04-15   22:01:07  &    17278  \\ 
		60002023035  &  2013-04-16   22:21:07  &    20279  \\ 
		60002023037  &  2013-04-18   00:16:07  &    17795  \\ 
		60002023039  &  2013-04-19   00:31:07  &    15958  \\
		\hline
	\end{tabular}
	\label{nupoint}
\end{table*}

\begin{table*}
	\centering
	\caption{Details of Swift pointings. The exposure times are rounded off to the nearest  seconds}
	\begin{tabular}{ c    c    c    c    c  }
		\hline
		\hline
obsid        &  start date and time   &  xrt exposure  &  uvot exposure  &  bat exposure  \\ \hline  
00035014061  &  2013-04-10  02:04:58  &    1079  &     1066  &    1085  \\  
00080050016  &  2013-04-11  00:30:59  &    1118  &     1076  &    1128  \\  
00032792001  &  2013-04-11  03:41:30  &    3488  &     3468  &    3502  \\  
00080050017  &  2013-04-11  21:48:59  &    1449  &     1419  &    1453  \\  
00080050018  &  2013-04-12  00:33:59  &    8726  &     8635  &    8746  \\  
00080050019  &  2013-04-12  21:53:58  &    9546  &     9428  &    9572  \\  
00032792002  &  2013-04-14  00:38:59  &    6327  &     6253  &    6362  \\  
00035014063  &  2013-04-14  23:04:59  &    4942  &     4874  &    4965  \\  
00035014062  &  2013-04-15  23:07:59  &     534  &      522  &     540  \\  
00035014064  &  2013-04-16  00:43:59  &   10262  &    10108  &   10302  \\  
00035014065  &  2013-04-17  00:46:59  &    8842  &     8731  &    8857  \\  
00035014066  &  2013-04-18  00:49:59  &    6887  &     6798  &    6907  \\ 
00035014067  &  2013-04-19  00:52:59  &    6132  &     6060  &    6152  \\  
00035014068  &  2013-04-20  00:55:59  &    5543  &     5482  &    4640  \\  
00035014069  &  2013-04-21  07:33:59  &     394  &      389  &     397  \\  
		\hline
	\end{tabular}
	\label{swiftpoint}
\end{table*}

\begin{table*}
	\centering
\caption{Reduced $\chi^2$ values for the various time bins (f1 to f13), as marked in Figure 3,  for the different models of the photon spectrum (broken power-law (bknpo) and log parabola (logpar)) and the particle spectrum (cutoff power law, broken power-law (bknpo) and logparabola) respectively. Observation ids of XRT and Nustar are also given. }
	\begin{tabular}{l | cc | cc | ccc}
		\hline
		\hline
		State & \multicolumn{2}{| c |}{Obs id} & \multicolumn{2}{| c |}{Photon spectrum} & \multicolumn{3}{| c }{Particle spectrum} \\ 
		&		XRT 			& 	Nustar           & 	bknpo  &  logpar  & CPL & BPL & LP 	  \\ \hline 
f1 		&	   00080050016 		&	60002023024      & 	1.27   & 	1.02  &		1.07	  &   	1.06	  &   	1.03   \\ 
f2 		& 	   00032792001		&	60002023025 (a)  & 	1.40   & 	1.16  &     1.26	  &   	1.19	  &   	1.16   \\ 
f3   	& 		- 				&   60002023025 (b)  &  0.97   &    0.92  &     1.01      &     1.06      &     1.01   \\ 
f4 		& 	   00080050017		& 	60002023025 (c)  & 	1.22   & 	1.04  &     1.08	  &   	1.07	  &   	1.05   \\ 
f5 		&	   00080050018		&	60002023025 (d)  & 	2.33   & 	1.05  &     1.21	  &   	1.16	  &   	1.06   \\ 
f6 		& 	   00080050019	 	&	60002023027      & 	2.93   & 	1.08  &     1.41	  &   	1.39	  &   	1.09   \\ 
f7 		& 	   00032792002	 	&	60002023029      & 	2.12   & 	1.10  &     1.21	  &   	1.19	  &   	1.10   \\ 
f8		& 	   00035014063	 	&	60002023031      & 	2.45   & 	1.14  &     1.20	  &   	1.31	  &   	1.13   \\ 
f9 		&	   00035014064	 	&	60002023033      & 	1.78   & 	0.93  &     1.06	  &   	1.09	  &   	0.93   \\ 
f10 	& 	   00035014065	 	&	60002023035      & 	1.75   & 	1.10  &     1.09	  &   	1.10	  &   	1.10   \\ 
f11 	& 	   00035014066	 	&	60002023037      & 	1.23   & 	1.00  &     1.03	  &   	1.06	  &   	1.00   \\ 
f12 	& 	   00035014067	 	&	60002023039      & 	1.14   & 	1.01  &     1.08	  &   	1.02	  &   	1.01   \\ 
f13  	&      00035014068      &   -                & 	1.07   &  	1.07  &     1.04      &     1.07      &     0.92   \\ 
		\hline
\end{tabular}
\label{jointid}
\end{table*}

\begin{table*}
	\centering
	\small
	\caption{Best fit parameters of the log parabolic photon and particle spectrum respectively.}
	\begin{tabular} {l | ccc | ccc }
		\hline
		\hline
		 & \multicolumn{3}{| c |}{Photon spectrum} & \multicolumn{3}{| c }{Particle spectrum} \\ 
		state& 	$\alpha_s$ 				&    $\beta_s$  		& 	$E_{p,s}$ 		  & $\alpha_p$        & $\beta_p$        & $E_{p,p}$\\	 \hline
f1 		& 	2.21	$\pm$	 0.02   & 	0.39  $\pm$ 0.01	& 	0.534	$\pm$ 	0.027	&	$3.19 \pm	0.06$ &	$1.96 \pm	0.11$ &	$0.50	\pm0.04 $    \\ 
f2 		& 	2.21	$\pm$	 0.01	& 	0.39  $\pm$ 0.01	& 	0.543	$\pm$ 	0.023	&	$3.18 \pm	0.04$ &	$1.98 \pm	0.09$ &	$0.50	\pm0.03 $    \\ 
f3 		& 	2.14    $\pm$	 0.05 	& 	0.27  $\pm$ 0.03    & 	0.550  $\pm$ 	0.074 	&	$3.11 \pm   0.14$ & $1.25 \pm	0.16$ & $0.36   \pm0.06	$ 	 \\  	
f4 		& 	1.92	$\pm$	 0.01	& 	0.38  $\pm$ 0.02	& 	1.254  $\pm$   0.0630 	&	$2.45 \pm	0.06$ &	$1.97 \pm	0.11$ &	$0.77   \pm0.06	$ 	 \\ 
f5 		& 	1.82	$\pm$	 0.01	& 	0.44  $\pm$ 0.01	& 	1.585  $\pm$   0.0229 	&	$2.10 \pm	0.02$ &	$2.35 \pm	0.04$ &	$0.95   \pm0.03	$    \\ 
f6 		& 	1.80	$\pm$	 0.01	& 	0.46  $\pm$ 0.01	& 	1.634  $\pm$   0.0205	&	$2.01 \pm	0.02$ &	$2.50 \pm	0.03$ &	$1.00   \pm0.02	$	 \\ 
f7 		& 	2.08	$\pm$	 0.01	& 	0.41  $\pm$ 0.01	& 	0.805	$\pm$   0.015	&	$2.82 \pm	0.03$ &	$2.10 \pm	0.04$ &	$0.64   \pm0.02	$ 	 \\ 
f8 		& 	1.61	$\pm$	 0.01	& 	0.37  $\pm$ 0.01	& 	3.272  $\pm$   0.0668	&	$1.66 \pm	0.03$ &	$1.92 \pm	0.04$ &	$1.23   \pm0.05 $    \\ 
f9 		& 	1.99	$\pm$	 0.01	& 	0.33  $\pm$ 0.01	& 	1.018  $\pm$   0.0201 	&	$2.69 \pm	0.02$ &	$1.62 \pm	0.03$ &	$0.61   \pm0.02 $ 	 \\ 
f10 	& 	1.83	$\pm$	 0.01	& 	0.30  $\pm$ 0.01	& 	1.872  $\pm$	0.039	 &	$2.33 \pm	0.02$ &	$1.46 \pm	0.03$ &	$0.77   \pm0.02 $    \\ 
f11 	& 	2.33	$\pm$	 0.01	&	0.30  $\pm$ 0.01	& 	0.280	$\pm$   0.013 	&   $3.55 \pm	0.04$ &	$1.44 \pm	0.07$ &	$0.29   \pm0.02	$    \\ 
f12 	& 	2.39	$\pm$	 0.01	& 	0.33  $\pm$ 0.01	& 	0.259	$\pm$   0.012	 &	$3.67 \pm	0.04$ &	$1.59 \pm	0.08$ &	$0.30   \pm0.02	$    \\ 
f13 	& 	2.25	$\pm$	 0.01	& 	0.35  $\pm$ 0.03	& 	0.445 $\pm$ 	0.049 	&   $3.30 \pm	0.05$ &	$1.78 \pm	0.23$ &	$0.43   \pm0.06 $    \\ \hline
	\end{tabular}
	\label{lp_par}
\end{table*}


\end{document}